
\input harvmac
\def\ha{\hbox to6.1pt{\hfil${\scriptstyle{1\over 2}}$\hfil}}
\def\pha{\hbox to5.5pt{\hfil${\scriptstyle{+}}$\hfil}}
\def\mha{\hbox to5.5pt{\hfil${\scriptstyle{-}}$\hfil}}
\def\znn{Z^{NS}_{NS}}
\def\zrn{Z^{R}_{NS}}
\def\znr{Z^{NS}_{R}}
\def\zrr{Z^{R}_{R}}
\def\tto{$SU(3)_c$$\times$$SU(2)_L$$\times$$U(1)_Y$}
\def\xmu{X^{\mu}}
\def\pmu{\psi ^{\mu}}
\def\ap{\alpha ^{\prime}}
\def\li{\lambda ^{i}}
\def\tach{|0,p^{\mu}\rangle}


\vglue .5in
\rightline{\hbox{hep-ph/yymmddd}}
\vglue .5in
\centerline{
\bf FOUR-DIMENSIONAL SUPERSTRING MODELS\footnote{$^\dagger$}
{Lectures at the 1995 Trieste Summer School
in High Energy Physics and Cosmology,
ICTP Trieste, 12 June - 28 July 1995.}}
\vskip.4in
\centerline{JOSEPH D. LYKKEN}
\bigskip\centerline{\it Fermi National Accelerator Laboratory}
\centerline{\it P.O. Box 500}
\centerline{\it Batavia, IL 60510}
\vskip 0.3in
\centerline{ABSTRACT}
\vskip 0.2in\noindent
These five lectures give an elementary introduction to
perturbative superstring theory,
superstring phenomenology, and the fermionic
construction of perturbative string models.
These lectures assume no prior knowledge of string theory.



\noindent
\newsec{Introduction and Outline}

%

Superstring theory is a unified description of gravity, gauge bosons,
and chiral matter. Superstring theory is free of ultraviolet infinities.
It is also exactly solvable (at least) in
perturbation theory around a
large class of backgrounds. Very recent developments seem to indicate
that there is only one consistent superstring theory,
which appears in
many avatars.

Superstring pheneomenology is the study of how superstring theory
makes contact with physics at accessible energy and length scales.
This field is still in its infancy: our idea of what
constitutes superstring phenomenology has evolved remarkably in the
past ten years, and is likely to be revised in profound ways as
we get a better handle on nonperturbative string effects.

There are {\it no} quantitative predictions, as yet, from
superstring theory. There are however a number of important
qualitative predictions and insights. Some of these are listed below.

\noindent 1.\ The existence of new particles: a dilaton, axion,
and perhaps other scalar moduli.

\noindent 2.\ Gauge coupling unification should occur, at a scale
not wildly different from the Planck scale, independently of
whether or not there is grand unification.

\noindent 3.\ Global continuous symmetries, in the effective
low energy theory that we see, are ``accidental''.

\noindent 4.\ Hidden sectors appear naturally.
This makes hidden sector dynamical supersymmetry breaking
seem more appealing. New SUSY breaking scenarios involving the
dilaton and other moduli fields are also possible.

\noindent 5.\ Gauged discrete and continuous flavor symmetries
appear naturally. This makes various schemes
for explaining the fermion mass and mixing hierarchies
seem more appealing.

\noindent 6.\ Having precisely three light generations of
standard model fermions is somehow special in string theory.
Furthermore the number of generations in a particular
string solution is correlated with all of the other low
energy properties of that solution. Thus in superstring
theory we can hope to eventually relate the number of generations
to, e.g., other features of the standard model.

This incomplete list is rather impressive, I would say.
However one does have to worry about the robustness of
even these qualitative statements, given our still rather
poor understanding of the true nature of string theory.
Many string theorists would advance the notion that
it is premature to even think about superstring phenomenology
in any serious way, either because it will turn out that
our current perturbative methods are a poor approximation
of anything, or because we are on the verge of solving
string theory completely in one fell swoop.

I have a different view. My guess is that we
are on the verge of achieving with string theory
the situation we currently have with respect to QCD.
For QCD we have a powerful qualitative understanding
of nonperturbative effects, via instantons, flux tubes,
monopole condensates, etc. (I will ignore the lattice in
making this analogy); this gives us confidence that we
understand the basic nature of the theory from first principles.
To compare numbers with experiment what we then do is to
parametrize our ignorance, using structure functions, etc.,
apply a few tricks like heavy quark symmetry, then do a
really good job on the perturbation theory. I expect that
current developments in string theory will similarly give
us confidence that we understand the basic nature of the
theory, and furthermore allow us to at least parametrize
our ignorance of nonperturbative effects. Then it will be
crucial to do a very good and thorough job on string
perturbation theory, supplemented by some ``tricks'' that
allow us to include certain nonperturbative information
in a controlled fashion.

If this scenario is correct we should be able to extract
a wealth of detailed information and insights from
superstring phenomenology, even though we cannot
compute from first principles how the string determines
its own vacuum state.

The outline of these lectures is as follows:

\leftline{LECTURE 1:}
\leftline{-- Perturbative string theory}

\leftline{LECTURE 2:}
\leftline{-- Supersymmetry on the worldsheet}
\leftline{-- The heterotic superstring in 10 dimensions and
4 dimensions}

\leftline{LECTURE 3:}
\leftline{-- Gauge coupling unification}
\leftline{-- Supersymmetry breaking}
\leftline{-- Fermion masses, fractional charge, proton decay}

\leftline{LECTURE 4:}
\leftline{-- Introduction to the fermionic construction
of four-dimensional superstring vacua}

\leftline{LECTURE 5:}
\leftline{-- A semi-realistic example: flipped $SU(5)$}
\leftline{-- A three generation
$SU(5)$ GUT}

\newsec{Perturbative String Theory}

We do not yet know what string theory really is. This is
because we do not have an adequate nonperturbative formulation
of the theory. What we have at the moment is a first-quantized
perturbative description of string theory as dynamics on
two-dimensional world-sheets. From a spacetime point of
view, perturbative string theory is simply a prescription
for computing S-matrix elements in a theory with
a finite number of massless particles and an infinite
number of massive particles.

In perturbative string theory the world-sheet dynamics
must involve a set of bosonic fields $\xmu (\sigma ^0,\sigma ^1)$,
where $\sigma ^0$, $\sigma ^1$ are the world-sheet proper
time and string coordinate, respectively. The superscript
$\mu$ is a spacetime vector index,
$\mu$$=$$0$,$1$,$\ldots$$D$$-$$1$; these bosonic fields are
thus a mapping from the world-sheet into a target space which
is some $D$-dimensional spacetime. These fields are necessary
to achieve Poincar\'e invariance of the spacetime
amplitudes. A priori there may be other world-sheet fields
as well.

We will be interested in closed strings, and thus apply
periodic boundary conditions to the $\xmu $:

\eqn\closebc{
\xmu (\sigma ^0,0) = \xmu (\sigma ^0,\pi)
}

What should we write down as a world-sheet action for
the $\xmu $? The simplest assumption is that they are
free fields. We will also assume that the world-sheet
action should be invariant under two-dimensional
general coordinate transformations; it is hard to imagine
a sensible theory otherwise. These two considerations
already suggest the form of the action:

\eqn\polact{
S  = {1\over 4\pi \ap} \int_{\Lambda_g} d^2\sigma\;
\sqrt{-{\rm det}g_{\alpha\beta}}\,g^{\alpha\beta}\,
\partial _{\alpha}X^{\mu}\partial _{\beta}X^{\nu}
\eta_{\mu\nu}
}
Here $\alpha$, $\beta = 0$, $1$ are worldsheet indices,
while $\mu$, $\nu = 0$,$\ldots$D$-$$1$ are spacetime
indices. $\eta _{\mu\nu}$ is the flat Minkowski metric
in spacetime, while $g_{\alpha\beta}$ is a world-sheet
metric. $\Lambda _g$ represents a two-dimensional surface
of genus $g$. $1/(2\pi\ap )$ is the string tension; the
parameter $\ap$ has units of length squared.

The world-sheet metric $g_{\alpha\beta}$ is a world-sheet
field, but it is an auxiliary field with no dynamics.
We could attempt to introduce some dynamics for
$g_{\alpha\beta}$ by adding to the action above an
Einstein-Hilbert term. However in two dimensions this
is a topological term:

\eqn\eh{
\int d^2\sigma \sqrt{-{\rm det}g}\; R = 4\pi\chi
}
where $\chi$ is the Euler characteristic of $\Lambda _g$.

Since $g_{\alpha\beta}$ is an auxiliary field we can, if we wish,
integrate it out. This then gives the Nambu-Goto form of the
action:

\eqn\nambu{
S = -{1\over 2\pi\ap} \big({\rm area\ of\ the\ worldsheet}\big)
}
which should not be surprising since the area of world-sheet
is the obvious nontrivial invariant under world-sheet general
coordinate transformations.

The action \polact\ has another local symmetry in addition
to two-dimensional general coordinate invariance. This
is Weyl invariance under a coordinate-dependent overall
rescaling of the world-sheet metric:

\eqn\weyl{
g_{\alpha\beta}(\sigma ^0,\sigma ^1) \to {\rm e}^{f(\sigma ^0,
\sigma ^1)}\; g_{\alpha\beta}(\sigma ^0,\sigma ^1)
}
We thus have three local world-sheet symmetries to
gauge-fix (two coordinate transformations + Weyl rescaling).
We can therefore fix three independent components of
$g_{\alpha\beta}$. For example, let us us go over to
complex coordinates:

\eqn\comco{
z = \sigma ^0 + i\sigma ^1\quad ;\quad \bar z = \sigma ^0 - i\sigma ^1
}
then we may write:

\eqn\liou{
\eqalign{
g_{zz} &= g_{\bar z\bar z} = 0\cr
g_{z\bar z} &= g_{\bar z z} = {1\over 2}{\rm e}^{\phi (z,\bar z)}\cr
}}
and the Liouville mode $\phi (z,\bar z)$ drops out of the action,
classically, due to Weyl invariance.

There is a large residual coordinate invariance after this
gauge-fixing. This is invariance under conformal transformations,
i.e., two-dimensional conformal mappings:

\eqn\map{
z \to f(z)\quad ;\quad \bar z \to f(\bar z)
}
Thus the ``flat-gauge'' action is a two-dimensional
conformal field theory of $D$ free bosons:

\eqn\flat{
S  = {1\over 4\pi \ap} \int_{\Lambda_g} dzd\bar z\;
\partial _{z}X^{\mu}\partial _{\bar z}X_{\mu}
}
The equations of motion

\eqn\eofmot{
\partial _z\partial _{\bar z}X^{\mu}(z,\bar z) = 0
}
indicate that $\partial _z\xmu $ is an analytic function of $z$,
while $\partial _{\bar z}\xmu $ is an antianalytic function, i.e.,
a function of $\bar z$. We often say that $\xmu$ splits into
separate left-moving and right-moving degrees of freedom.

If we now define normal mode operators $\alpha ^{\mu}_n$,
${\tilde \alpha}^{\mu}_n$ by

\eqn\normode{
\eqalign{
\partial _z\xmu &\sim \sum_n z^{-n-1}\,\alpha ^{\mu}_n \cr
\partial _{\bar z}\xmu
&\sim \sum_n {\bar z}^{-n-1}\,{\tilde \alpha}^{\mu}_n \cr
}}
we can then define a Hilbert space of states.

Let us pause for a moment. At this point what we are calling
perturbative string theory seems absolutely trivial, and
furthermore has no apparent relation to real physics.
These objections will disappear once we quantize the theory.
To properly quantize a theory with local symmetries, we
must either introduce Fadeev-Popov ghosts on the the
world-sheet, or perform a canonical quantization with
constraints on the physical states.
Let us sketch the canonical approach.

The three local world-sheet symmetries imply three
independent first-class constraints, namely, that
the three independent components of the
two-dimensional symmetric
energy-momentum tensor $T_{ab}$ must annihilate all
physical states. Actually, since $T_{z\bar z}$
vanishes by the equation of motion, we only have to worry
about two contraint operators:

\eqn\emops{
\eqalign{
T(z) &\equiv T_{zz} = -{1\over 2}\partial _z\xmu\partial _zX_{\mu}\cr
\bar T(\bar z) &\equiv T_{\bar z \bar z}
= -{1\over 2}\partial _{\bar z}\xmu\partial _{\bar z}X_{\mu}\cr
}}
A mode expansion of these defines the Virasoro mode operators
$L_n$, ${\bar L}_n$:
\eqn\virmode{
T(z) = \sum_n z^{-n-2}\,L_n
}

Upon canonically quantizing this system, we encounter a
surprise: the commutator [$T(z)$,$T(z^{\prime})$] has a
Schwinger term, indicative of a possible anomaly. In terms
of modes
\eqn\viralg{
\left[ L_n,L_m \right] = (n-m)L_{n+m} + {c\over 12}n(n^2-1)\delta_{n,-m}
}
This Virasoro algebra is, more generally, always a subalgebra of
the constraint algebra for any two-dimensional conformal field
theory. The second term is the Schwinger or ``central'' term;
the coefficient $c$, which equals $D$ in the case at hand, is
called the central charge.

Using \normode ,\emops , and \virmode\ we can express the
Virasoro constraint operators in terms of normal mode
operators, up to a normal ordering ambiguity in the definition
of $L_0$. We thus write the constraints on physical states as
\eqn\phycon{
\eqalign{
L_n |{\rm phys}\rangle &= 0\quad ;\quad {\rm for\ all\ }n>0 \cr
\left( L_0 - a \right) |{\rm phys}\rangle &= 0\cr
{\bar L}_n |{\rm phys}\rangle &= 0\quad ;
\quad {\rm for\ all\ }n>0 \cr
\left( {\bar L}_0 - a \right) |{\rm phys}\rangle &= 0\cr
\left( L_0 - {\bar L}_0 \right) |{\rm phys}\rangle &= 0\cr
}}
where $a$ represents the normal ordering ambiguity.

To get the full spectrum of physical states, one
would then fix the residual conformal symmetry by the
light-cone gauge condition:
$X^+(\sigma ^0,\sigma ^1)$$\propto$$\sigma^0$. This leaves
only the $D$$-$$2$ transverse components of $X$. The physical
states are then all states which can be constructed using the
transverse mode operators $\alpha _n^i$, ${\tilde \alpha}_n^i$,
and satisfying the constraints \phycon . This is clearly not
a trivial theory.

For general values of $D$ the physical state spectrum will
turn out to contain states with negative norm. This is an
indication that the theory is not unitary, due to the presence
of an anomaly. The anomaly in this case is in the Weyl symmetry:
the Liouville mode which decouples in the classical action
does not in general decouple in the quantum theory.

Fortunately for precisely the values $D$$=$$26$, $a$$=$$1$, all
is well: the spectrum is unitary and the anomaly absent.
We say that $D$$=$$26$ is the critical dimension which
defines a consistent bosonic string theory. One can also
see from the spacetime point of view that precisely for
these values the theory respects Lorentz invariance at the
quantum level.

Let us now compute the mass and spin of some of the
low-lying physical states. We note first that spacetime
four-momentum for string states is defined from the
center of mass momentum of the string. This in turn
corresponds to the zero mode operators
\eqn\zerom{
\alpha_0^i = {\tilde \alpha}_0^i = \sqrt{{\ap\over 2}}\,p^i
}
where the superscript $i$ denotes transverse components.

We then recognize the constraint equations in \phycon\
involving $L_0$, ${\bar L}_0$ as mass-shell conditions:
\eqn\mshell{
\eqalign{
\ap M^2 |{\rm phys}\rangle
&\equiv -\ap p^{\mu}p_{\mu}|{\rm phys}\rangle\cr
&= \left[ 2(N-1) + 2(\tilde N -1) \right] |{\rm phys}\rangle\cr
}}
where $N$, $\tilde N$ are occupation numbers:
\eqn\nguys{
\eqalign{
N &= \sum_{n=1}^{\infty} \alpha_{-n}^i \alpha _n^i \cr
\tilde N &=
\sum_{n=1}^{\infty} {\tilde \alpha}_{-n}^i {\tilde \alpha}_n^i \cr
(N &- \tilde N) |{\rm phys}\rangle = 0\cr
}}

The simplest physical state has $N$$=$$\tilde N$$=$$0$; let
us call this state $\tach$. We may think of $\tach$ as being
created from the conformal field theory vacuum by a vertex
operator:
\eqn\tvop{
\tach = : {\rm e}^{ip_{\mu}X^{\mu}(0,0)} : |0\rangle
}
where ``::'' denotes normal ordering. The vertex operator
is a local operator which creates a physical state at a
point on the world-sheet.

The mass-shell condition
on this state is:
\eqn\tmass{
\ap M^2 \tach = -4\tach
}
indicating that this state represents an on-shell
scalar tachyon!

This is a little alarming, but let us continue. The next
simplest physical state has $N$$=$$\tilde N$$=$$1$.
We can express this state in terms of normal mode
operators acting on $\tach$:
\eqn\grav{
|\Omega^{ij},p^{\mu}\rangle =
\alpha_{-1}^i{\tilde \alpha}_{-1}^j
\tach
}
and the mass-shell condition is:
\eqn\mgrav{
\ap M^2 |\Omega^{ij},p^{\mu}\rangle = 0
}
Thus we have $(D$$-$$2)^2$ on-shell massless states.
The states corresponding to the symmetric part of
the second rank tensor $\Omega ^{ij}$ are the transverse
components of a massless spin 2 particle, i.e., a graviton!
The trace part is a massless scalar ``dilaton'', and the
antisymmetric parts represent a massless antisymmetric
tensor field (in $D$$=$$4$ this would just be a massless
pseudoscalar ``axion'').

Thus the bosonic string is a quantum theory of gravity plus matter.
To be consistent we should fix $\ap$ in terms of the Planck
scale (or equivalently Newton's constant):
\eqn\mplap{
\ap  \sim {1\over M_{pl}^2}
}

\subsec{Green Functions in String Theory}

As already indicated, we use vertex operators to create
on-shell physical states at points on worldsheet surfaces.
Thus the external states of an on-shell n-point function
are represented by n vertex operator insertions at
n points on a closed two-dimensional surface. Since the
world-sheet metric was gauged away, the only distinct
two-dimensional surfaces are those differing either
topologically or by some other {\it global} parameters.
The topology of closed two-dimensional manifolds
is completely classified as the Riemann surfaces of
genus $g$. Genus $g$$=$$0$ is the sphere, $g$$=$$1$ is the
torus, $g$$=$$2$ is the double torus, etc..

An n-point function is then {\it defined} to be a
simple sum of path-integrals over the distinct
world-sheet surfaces, with vertex operator insertions
at n points $z_i$, and integrals over the locations of
these points:
\eqn\npf{
\sum _g {\rm e}^{(n+2g-2)\phi_0}\int [dXdg]\,{\rm e}^{-S}\,
\prod ^n_{i=1} \int d^2z_i\; V_i(z_i)
}
The first exponential occurs because we have been
careful to include an Einstein-Hilbert term in the
action with coefficient $\phi _0/4\pi$; using \eh\
we then obtain the exponential by using the
following expression for the Euler characteristic
of a Riemann surface punctured at n points:
\eqn\eulc{
\chi = 2 - 2g - n
}

It is possible to show that the parameter $\phi _0$
has a physical interpretation: it is the vacuum
expectation value (vev) of the dilaton field.
We note further that the sum over genus in \npf\
can be interpreted as a perturbation series in
an effective coupling constant exp($2\phi_0$).
We therefore define the string coupling $g_s$ by
\eqn\strcoup{
{g_s^2\over 4\pi} \equiv {\rm e}^{2\phi_0}
}

This is what we mean by saying that string theory
defined as conformal field theory on two-dimensional
closed surfaces is really just {\it perturbative}
string theory.

Note that in \npf\ we denoted an integral over world-sheet
metrics, despite the fact that this metric can be gauged
away. This is because the world-sheet metric can really
only be gauged away {\it locally}; for genus 1 or larger
there are in fact additional global parameters which we
must integrate over appropriate domains.
For example with the torus there
is a complex parameter $\tau$ which distinguishes
inequivalent tori. We want two-dimensional
general coordinate invariance to apply with respect
to these global parameters -- this is referred to as
``modular invariance''. Thus
we need to impose additional restrictions
on the quantum theory to ensure modular invariant Green
functions. Although modular invariance sounds rather
abstract, it is related to two very physical properties of
tree-level string theory, namely, s-t-u duality of the
tree-level 4-point functions, and the fact that the vertex
operators form a mutually local closed associative algebra.

\newsec{Superstrings}

\subsec{Supersymmetry on the world-sheet}

The spectrum of the
$D$$=$$26$ perturbative closed bosonic string contains
a tachyon. This is not a satisfactory situation.
Can we reformulate perturbative string theory in
such a way as to remove the tachyon while keeping
the graviton?

One way to do this is to incorporate spacetime
supersymmetry into string theory (recall that we
have already implemented spacetime Poincar\'e invariance).
If the string spectrum were spacetime supersymmetric,
the scalar tachyon would have a superpartner -- a
fermionic tachyon. However since the Dirac equation does
not possess tachyonic solutions, what should happen
instead is that the tachyon is simply removed from
the physical spectrum by additional physical state
constraints.

As far as we know, the only way to implement
spacetime supersymmetry in string theory is
to first implement {\it world-sheet} supersymmetry.
This is rather straightforward to accomplish.
Suppose we want $N$$=$$1$ world-sheet supersymmetry;
clearly we should introduce superpartners for the $\xmu$:
\eqn\xsp{
\xmu \Rightarrow \xmu,\, \pmu
}
where $\pmu (z,\bar z)$ are $D$ Majorana world-sheet
fermions. Note that these {\it world-sheet fermions}
carry a {\it spacetime vector} index $\mu$.

It is easy to see how to modify the flat-gauge action
\flat\ :
\eqn\sflat{
\int d^2\sigma \left[ \partial _{\alpha}\xmu
\partial ^{\alpha}X_{\mu} - i {\bar \psi}^{\mu}\gamma ^{\alpha}
\partial _{\alpha} \pmu \right]
}
where the $\gamma ^{\alpha}$ are two-dimensional
gamma matrices. This gauge-fixed action now exhibits
{\it global} world-sheet SUSY; with more work one
can improve the ungauge-fixed action \polact\ to
exhibit {\it local} world-sheet SUSY.

The $\pmu$ contribute to $T(z)$, ${\bar T}(\bar z)$ and thus to
the central charge:
\eqn\nccharge{
c = D + {D\over 2} = {3D\over 2}
}

In addition, local worldsheet SUSY implies new constraint
operators (the superpartners of $T(z)$, ${\bar T}(\bar z)$):
$T_F(z)$, ${\bar T}_F(\bar z)$. These promote the Virasoro
algebra to an $N$$=$$1$ superconformal algebra.
The explicit form of these constraint operators is
\eqn\formoft{
\eqalign{
T_F(z) &\sim \partial _z\xmu \psi _{\mu}\cr
{\bar T}_F(\bar z) &\sim \partial _{\bar z}\xmu {\bar \psi}_{\mu}\cr
}}
where we have broken up the Majorana fermions into
their left and right-moving Majorana-Weyl components:
\eqn\mwcomp{
\psi ^{\mu}(z,\bar z) =
\left( {\pmu (z)\atop {\bar \psi}^{\mu}(\bar z)} \right)
}

Because we have added new degrees
of freedom and new local symmetries on the
world-sheet, we must reexamine the question of the
Weyl anomaly. This is easier to discuss in the
language of covariant quantization, where the
condition for anomaly freedom is that the total central
charge, {\it including} contributions from world-sheet
ghosts, should vanish. For the original bosonic
string the world-sheet ghosts associated with
conformal invariance contribute central charge
$c$$=$$-$$26$, hence the statement that the critical
dimension for $D$ is $26$. When we promote conformal
invariance to superconformal invariance, we must
introduce another set of ghosts; the total ghost
contribution to the central charge then becomes
$-15$. Each Majorana worldsheet fermion contributes
$c$$=$$1/2$; thus anomaly freedom requires:
\eqn\noamom{
D + {D\over 2} - 15 = 0
}
Thus $D$$=$$10$ is the critical dimension of the
superstring.

Also, the normal ordering constant in the mass-shell
conditions changes:
\eqn\newms{
\left( L_0 - 1 \right) |{\rm phys}\rangle = 0
\Rightarrow
\left( L_0 - \half \right) |{\rm phys}\rangle = 0
}
Notice, however, that there is still the danger
of obtaining a tachyon in the physical spectrum.

\subsec{Boundary Conditions}

We imposed periodic boundary conditions for the
$\xmu$; for the Majorana fermions $\pmu$ we may
choose either periodic or antiperiodic. We will
refer to world-sheet fermions with periodic
boundary conditions
as Ramond or R, and refer to fermions with
antiperiodic boundary conditions as
Neveu-Schwarz or NS. The R/NS notation is
superior because the notion of periodic/antiperiodic
can be interchanged by a conformal mapping.

What is the difference between R and NS boundary
conditions from the point of view of two-dimensional
conformal field theory? There is a unique conformal
field theory vacuum state $|0\rangle$; this can be
considered the Fock vacuum of the NS modes of the
worldsheet Majorana-Weyl fermions $\pmu (z)$,
${\bar \psi}^{\mu}(\bar z)$. The ``Neveu-Schwarz
sector'' of the conformal field theory corresponds
to the states built upon this vacuum with the NS modes.

The conformal field theory also contains
$4$$*$$D$ additional local operators called
{\it twist fields}: one set
$\sigma (z)$, $\mu (z)$,
$\bar \sigma (\bar z)$, $\bar \mu (\bar z)$,
for each $\pmu$, ${\bar \psi}^{\mu}$.
These operators are necessary for the
consistency of the conformal field theory.
(One might reasonably ask why there are no twist fields
in the action, but answering that question
completely requires a fairly nasty detour.)
The states $\sigma (0)|0\rangle$, $\mu (0)|0\rangle$ define
a doubly degenerate left-moving {\it Ramond vacuum} (and
similarly for barred fields). The ``Ramond sector''
of the conformal field theory corresponds to the states
built upon the Ramond vacuum with the R modes of
$\pmu$, ${\bar \psi}^{\mu}$. Note that from the conformal
field theory point of view there is only one
type of Majorana fermion field, not two; the existence
of both NS and R modes coming from the same fermion field is a
result of the fact that the operator product of
a fermion field with a twist field is nonlocal.

\subsec{Spacetime Supersymmetry}

So far we have seen that the Hilbert space
of a superconformal field theory
factorizes into Neveu-Schwarz sectors
and Ramond sectors. We may also wonder what
other role the twist fields play in the theory.
For simplicity, suppose that $D$ is even and
that we have paired up the $D$$-$$2$ transverse left-moving
Majorana-Weyl world-sheet fermions $\pmu (z)$
to make $(D$$-$$2)/2$ left-moving Weyl fermions
(and similarly for the right-movers). There are
a pair of left-moving twist fields associated with
each left-moving Weyl fermion; thus the total
left-moving Ramond vacuum degeneracy is $2^{(D-2)/2}$.
Futhermore, we can define a local operator $S_{\alpha}(z)$,
called a {\it spin field}, which is the product of the
$(D$$-$$2)/2$ left-moving twist fields. The index $\alpha$
takes $2^{(D-2)/2}$ values; this can be interpreted as
{\it a spacetime spinor index}. So we see that the other
role of twist fields in the perturbative superstring
is that they allow us to construct states which are
spacetime fermions!

Now we have enough formalism to discuss
spacetime supersymmetry. In a first-quantized
formalism like perturbative string theory,
the presence of (local) spacetime SUSY is most
easily identified by looking for a massless
spin 3/2 particle -- a gravitino -- in the physical
spectrum. This means there must be vertex operator
which creates the gravitino. The form of this
vertex operator is:
\eqn\spsusy{
V_{\alpha}^{\mu}(z,\bar z) \sim
S_{\alpha}(z)\; \Sigma (z)\; \partial _{\bar z}\xmu \;
{\rm e}^{ip\cdot X}
}
where I have suppressed ghost field dependence, and
where $\Sigma (z)$ is an analytic field with some
known conformal field theory properties. In fact,
the existence of $\Sigma (z)$ in the conformal field
theory can be shown to imply the existence of
an abelian current, $J(z)$, which extends the local
$N$$=$$1$ superconformal world-sheet symmetry to
a global $N$$=$$2$ superconformal world-sheet symmetry:
\eqn\onetotwo{
T(z),\,T_F(z) \Rightarrow T(z),\, T_F^+(z),\, T_F^-(z),\,
J(z)
}
This is a beautiful example of an
explicit relation between spacetime and world-sheet
symmetries.

\subsec{GSO Projections}

So far we have discussed vertex operators for three
kinds of particles: the tachyon, the graviton, and
the gravitino. The states created by these vertex
operators do satisfy the superconformal contraint
equations for physical states. However we have
already mentioned the fact that the physical spectrum
must be consistent with additional constraints from
modular invariance. In particular we mentioned the
tree-level modular invariance constraint that
all of the physical vertex operators must be
mutually local. In general we have to perform
additional projections on the physical states
to some subset that obeys the additional
constraints. Such projections are known
as {\it GSO projections}.

The locality properties of the
operator products of our three vertex operators
are given below:
\eqn\locality{
\eqalign{
{\rm tachyon}*{\rm graviton} &: \;local\cr
{\rm graviton}*{\rm gravitino} &: \;local\cr
{\rm tachyon}*{\rm gravitino} &: \;nonlocal\cr
}}

Thus with regard to these three states, we see that
a GSO projection is indeed required. Furthermore
there are two possible choices of projection:
the first one removes the gravitino from the
physical spectrum, while the second choice removes
the tachyon from the physical spectrum. In the
first case we will end up with a superstring
whose spectrum contains a tachyon;
in the second case we end up with a {\it tachyon-free}
superstring which exhibits spacetime supersymmetry.

\subsec{The 10-dimensional Heterotic Superstring}

The name ``heterotic'' refers to the following
trick: as we have seen, the world-sheet Majorana fermions
$\pmu (z,\bar z)$ split up into separate left-moving
and right-moving Majorana-Weyl fermions
$\pmu (z)$, ${\bar \psi}^{\mu}(\bar z)$. Thus we can
in principle implement $N$$=$$1$ superconformal invariance
on, e.g., just the right-movers. In this case our
world-sheet degrees of freedom will be
\eqn\hetdeg{
\xmu (z,\bar z),\quad {\bar \psi}^{\mu}(\bar z);
\quad \mu = 0,\ldots 9
}
The right-moving contributions to the Weyl anomaly
will cancel, however for the left-moving part we need
additional central charge $26$$-$$10$$=$$16$ to
remove the anomaly. The simplest choice of
additional world-sheet fields to accomplish this
is 32 left-moving Majorana-Weyl free fermions:
$\li (z)$, $i$$=$$1,\ldots 32$.

The flat gauge action of the 10-dimensional
heterotic superstring is thus
\eqn\tenact{
\int d^2z \left[ \partial _z\xmu \partial _{\bar z}
X_{\mu} - 2i{\bar \psi}^{\mu}(\bar z)
\partial _{\bar z}{\bar \psi}_{\mu}(\bar z)
-2i\li (z)\partial _z\li (z) \right]
}

The $\li (z)$ do not carry a spacetime index;
they can be regarded as an ``internal'' part
of the conformal field theory. However
these new worldsheet fermions have an
important effect on the physical spectrum.
This is because the normal ordered product
\eqn\bilins{
:\li (z) \lambda ^j(z):
}
for any $i$$\ne$$j$ is a world-sheet current.
These 496 currents define an operator
product current algebra usually called
a Kac-Moody algebra:
\eqn\kmalg{
J^A(z_1)J^B(z_2) = {k\delta ^{AB}\over (z_1-z_2)^2}
+ {if^{ABC}J^C(z_2)\over (z_1-z_2)} + \;\ldots
}
where $f^{ABC}$ are structure constants of $SO(32)$,
and $k$ is the Kac-Moody level, which equals 1 in
this case. We have assumed all NS boundary conditions
for the $\li$ in this discussion; later we will see
that by being more careful with the NS and R sectors
we can also realize an $E_8$$\times$$E_8$ heterotic
superstring.

These currents allow us to construct vertex operators
that make gauge bosons:
\eqn\gbvert{
V^{\mu A}(z,\bar z) \sim {1\over\sqrt{k}} J^A(z)
{\bar \psi}^{\mu}(\bar z) {\rm e}^{ip\cdot X}
}
Thus the massless spectrum of the 10-dimensional
heterotic superstring consists of a graviton,
dilaton, antisymmetric tensor, gauge bosons of
$SO(32)$ or $E_8$$\times$$E_8$, plus superpartners for
all of the above.

\subsec{The 4-dimensional Heterotic Superstring}

Our construction of the 10-dimensional heterotic
superstring suggests an obvious 4-dimensional
counterpart. Suppose we take the same world-sheet
degrees of freedom as for the 10-dimensional
heterotic string, but let the spacetime index
$\mu$ run over only $0$--$3$. We then would require
additional
left-moving central charge 6, and right-moving
central charge 9, to cancel the Weyl anomaly.
We can do this very simply by adding 6 left-moving
and 9 right-moving free world-sheet Weyl fermions:
\eqn\newdfg{
\eqalign{
\chi ^a(z)&,\qquad a=1,\ldots 6\cr
{\bar \chi}^{\alpha}(\bar z)&,\qquad \alpha = 1,\ldots 9\cr
}}

If we align the boundary conditions of these new
Weyl fermions, we
can indeed construct a consistent modular invariant
4-dimensional heterotic superstring in this way.
Since we live in a 4-dimensional world, we may
therefore ask: who needs 10-dimensional superstrings?
The answer is that the 4-dimensional heterotic superstring
just described is not a new superstring -- it is the
precisely the same string expanded around a different
background. To be precise, what we called the 10-dimensional
heterotic superstring should more properly be called
the heterotic superstring expanded around 10-dimensional
Minkowski space. What we called the 4-dimensional
heterotic superstring should more properly be called
the heterotic superstring expanded around
4-dimensional Minkowski space $\times$ a 6-dimensional
torus whose radii (in string units) are all unity.
The fact that we can represent this torus at a certain
radius by Weyl fermions is an example of
the well-known phenomenon of {\it bosonization}
in two-dimensional field theories.

Our current belief is that there is only
one heterotic superstring (in fact, only one superstring).
Conformal field theory solutions can exhibit this
superstring as a perturbative expansion around
a huge variety of backgrounds (also called string vacua).
These different solutions are often referred to as
``compactifications'', although this terminology
is only meaningful when the world-sheet degrees
of freedom have an obvious geometrical interpretation,
which in general they don't.

\subsec{Moduli}

In the example above of the heterotic string
compactified on a six-torus, the radii of the torus
obviously define a continuously connected family of
(perturbatively) consistent string vacua. Since
all of these vacua respect spacetime supersymmetry,
they are exactly degenerate. From the point of view
of the effective spacetime field theory, there must
be scalar fields, called {\it moduli}, whose
potentials have flat directions. Turning on vevs
for these moduli fields then corresponds to
changing the radii of the torus --i.e., moving
around in a space of continuously connected string solutions,
called {\it moduli space}.

Obviously a key problem in string theory is to
map out and characterize the full moduli space.
The real world, perhaps, would correspond to
one point in that moduli space, but string
perturbation theory gives us no clue as to how
this huge degeneracy of degenerate vacua is
lifted, or even as to whether or not it {\it is} lifted.

The appearance of moduli with undetermined vevs is
a serious problem for superstring phenomenology.
Moduli vevs, like the Higgs vev in the standard model,
contribute to coupling constants in the low energy
effective field theory. Thus the low energy physics
is completely dependent on where we are in moduli
space. Furthermore the dilaton field is a modulus;
its vev, undetermined in string perturbation theory,
fixes the effective string coupling $g_s$
that defines the perturbative string expansion.
It is easy to find regions in moduli space where
the string is strongly coupled, but in such regions
our perturbative string apparatus is presumably
breaking down. One hopes that nonperturbative
effects will stabilize the dilaton vev and the
other moduli vevs without otherwise doing too
much damage to the picture provided by perturbative
calculations, but this is not at all obvious.

Fermionic string constructions (which is our focus
in these lectures) always sample special points
in moduli space, while more geometrical
constructions like orbifolds sample entire
regions of moduli space. Thus it is
useful to think about fermionic solutions in
terms of an equivalent orbifold description.
This is not always possible, however, as the
fermionic construction samples points in the
full moduli space which cannot be reached
by the orbifold construction.

\newsec{Superstring Phenomenology}

As long as we are restricted to perturbative string theory,
we probably do not have enough information to determine
if any string solution corresponds precisely to the real
world. The problem is compounded by our ignorance of
what ``the real world'' looks like at energy scales
close to the Planck scale.

Nevertheless there is much to be learned by examining
string solutions which share at least some of the
features of the standard model, and of the minimal
supersymmetric standard model (MSSM). For some
purposes it may be sufficient to look at perturbative
string vacua which share only one or two features of
the MSSM.
However it is clearly important to focus on models
which share {\it many} of the key features of the
MSSM; this is because in string theory {\it all}
phenomenological properties of the effective low energy
theory are {\it related} by world-sheet symmetries.
To understand superstring physics we need to understand
these relationships.

It is useful therefore to define the concept of a
``semi-realistic'' superstring model. I will define
this to be a perturbative string solution which
satisfies the following criteria:

\noindent -- Four uncompactified dimensions.

\noindent -- N=1 spacetime supersymmetry (perturbatively,
at the string scale).

\noindent -- three light generations of standard model
fermions.

\noindent -- the gauge group includes
\tto , or embeds it in a larger group like $SU(5)$
while providing some mechanism (e.g. Higgs) to break
the larger group.

These may seem like rather weak criteria for
``semi-realistic'', but they will suffice for
two reasons. The first is that, despite a decade
of effort, string theorists have
constructed only about two dozen (depending on how
you count)
superstring models that meet these criteria.
The second reason is that it turns out that
models which meet this criteria often
display a number of other desirable phenomenological
properties.

We will discuss next two such properties which
show up rather generally in semi-realistic string
models: gauge coupling unification, and mechanisms
for dynamical SUSY breaking.

\subsec{Gauge Coupling Unification}

Recall that perturbative string theory contains only
two fundamental parameters: $\ap$ which has units
of length squared, and the effective string coupling,
$g_s$, which is dimensionless. Suppose I now consider,
at string tree-level, the four-point amplitude for
on-shell gauge boson scattering. This amplitude is
of course proportional to $g^2/4\pi$, where $g$
is the appropriate gauge-coupling defined at some
scale. We can therefore relate this field theory
coupling to string parameters:
\eqn\gtos{
{g^2\over 4\pi} = {g_s^2\over 4\pi}(\ap )^3 {1\over V_6}
}
where $V_6$ is some function,
with dimension ${\rm (length)}^6$,
of the moduli vevs. In string models with a geometric
interpretation $V_6$ is roughly the volume
of the compactified space.

This relationship is of limited use since we
don't know how to compute either $g_s$ or $V_6$.
We can obtain a much cleaner relationship
by computing a {\it ratio}, i.e. the ratio
of the graviton exchange contribution to
this amplitude over the gauge boson exchange
contribution. Consider first graviton exchange
in, say, the s-channel, in the limit
as $s$$\to$$0$. There is a factor of
$\sqrt{8\pi}/M_{pl}$, field theoretically,
from each gauge-gauge-graviton vertex.
Thus in field theory this diagram goes like
$8\pi /M_{pl}^2$, compared to the gauge boson
exchange contribution which goes like $g^2$.
In string theory the ratio of these
contributions is just $k\ap$. The explicit
factor of the Kac-Moody level appears for
the following reason: in the graviton exchange
diagram we contract two pairs of currents
with Kronecker deltas, introducing (by \kmalg )
two factors of $k$; for the gauge boson exchange
diagram we contract two pairs of currents
to make two new currents times structure constants
(see the second term in \kmalg ), then contract those
currents to produce a single factor of $k$.

We can now equate the field theory ratio with the string
theory ratio (being careful that
we have not left out any numerical factors):
\eqn\gcur{
{16\pi\over g^2M_{pl}^2} = k\ap
}
Furthermore, this relation holds for all of the gauge
couplings! Thus, for example, for a string model which
contains \tto , we have the string tree-level relation:
\eqn\gcurel{
k_3g_3^2 = k_2g_2^2 = k_1g_1^2 = {16\pi\over \ap M_{pl}^2}
}
at some appropriate running scale called {\it the string scale}.
This scale, defined by minimizing threshold effects, turns
out to be approximately $5$$\times$$10^{17}$ GeV. We also see
from \gcurel\ that, for $kg^2$$\sim$$1$, the fundamental
dimensionful unit of string theory is approximately
$10^{18}$ GeV, in energy units.

We have thus discovered a very general prediction
of semi-realistic superstring models with
$k_3$$=$$k_2$$=$$k_1$: gauge coupling unification
occurs (modulo threshold corrections) at a scale
of roughly $5$$\times$$10^{17}$ GeV. This is a
remarkable prediction for two reasons. The first is
that no mention was made here of grand unification,
which is the only known method in field theory of
unifying gauge couplings. The second is that the
predicted unification scale, $5$$\times$$10^{17}$ GeV,
is only a factor of 10 - 20 larger than the scale
suggested by the MSSM and low energy data:
$3$$\times$$10^{16}$ GeV. Put another way, there is
less than 10\% disagreement in the exponents.

There has been much effort to improve this
prediction by explaining the remaining
discrepancy. Some possible explanations:

\noindent -- The Kac-Moody levels are not all equal.
The problem with this idea is that for nonabelian
groups the Kac-Moody levels are integers; thus it is
difficult to tune the ratio $k_3/k_2$ without going
to quite large levels (which then has other
unfortunate consequences). However $k_1$ is
adjustable, since for abelian groups the
``Kac-Moody level'' is really just a
model-dependent normalization factor.

\noindent -- The string threshold corrections
may be large. This corresponds to a modulus
vev getting a large value or perhaps several
moduli vevs getting moderate values. In the
semi-realistic models constructed so far this
doesn't appear to happen, but this is still an
attractive solution.

\noindent -- There is a GUT, e.g. $SU(5)$, which is
then broken dynamically or by adjoint Higgs at
$3$$\times$$10^{16}$ GeV. This seems a little ugly,
but does have the advantage of introducing a
small parameter into the low energy theory, namely,
the ratio of the GUT scale to the string scale.

\noindent -- There is exotic matter beyond the MSSM
content, which thus effects the running of the
standard model gauge couplings between $M_Z$ and
the string scale. This
possiblity is currently being investigated
in some known semi-realistic models. The
problem with this solution is ugliness;
in particular, the contribution of any
not-too-exotic extra matter on the gauge coupling
running is rather large, requiring rather large
cancelling effects to get the right result.

\subsec{Supersymmetry Breaking}

Supersymmetry must be broken, at an effective scale of
100 GeV - 1 TeV, to agree with the real world.
If the SUSY breaking sector of the theory is
``hidden'' from the sector that contains the MSSM
--e.g. they are coupled only gravitationally--
then the SUSY breaking scale can be very large
$\sim$$10^{13}$-$10^{14}$ GeV.
In string theory there are many possibilities for
how this happens:

\noindent -- SUSY breaking is field theoretic, i.e.,
in the effective supergravity field theory below
the string scale, SUSY is broken spontaneously
or dynamically. The scale is set either by moduli
vevs or by a gauge coupling getting strong.
SUSY breaking in field theory means
giving a vev to either the F auxiliary field
of a chiral superfield
\eqn\csupf{
\Phi = \varphi + \theta\psi + \theta\theta F
}
or to the D term of a vector superfield.

\noindent -- SUSY breaking is inherently stringy.
This can happen in string perturbation theory when
a GSO projection associated with keeping the radius
moduli of a compactification {\it in} the spectrum,
projects the gravitino {\it out} of the spectrum.
In this case the scale of SUSY breaking is related
to the compactification scale. However we see from
\gtos\ that the compactification scale is naturally
close to $10^{18}$ GeV when the string coupling is
reasonably weak -- which we have assumed is the case
in order to apply string perturbation theory! So
this does not seem to be a promising mechanism.
String nonperturbative effects may contribute to
SUSY breaking, but at the moment there isn't much
that we can say about this.

\subsec{Hidden Sector Gaugino Condensate}

Much of the work on SUSY breaking in string theory
has focused on the idea that SUSY breaking is
triggered by the formation of a gaugino
condensate in a hidden sector. If the hidden
sector contains a fairly large nonabelian group
and the hidden matter content is such that
the hidden gauge coupling is asymptotically
free, then gaugino condensation may occur at
a high scale $M$$\sim$$10^{13}$ GeV. The known
semi-realistic string models at least come
close to meeting these conditions, making this
scenario seem rather natural in perturbative
string theory. This should be regarded as a
nontrivial success of string theory, since
these ingredients are in no way built into
our definition of ``semi-realistic''.

However the hidden gaugino condensate
\eqn\hgcond{
\langle \lambda \lambda \rangle \sim M^3
}
does not itself break supersymmetry, because
$\lambda\lambda$ is not an F term. Thus gaugino
condensation is supposed to trigger SUSY breaking
indirectly through its effect on other fields:
either moduli fields (including the dilaton) or
hidden sector chiral matter.

In the visible sector we would see SUSY-breaking
only via gravitational effects, or more generally
via Planck-mass-suppressed effects. The effective
scale of SUSY breaking in the visible sector
is set by the gravitino mass:
\eqn\gravmassrel{
m_{3/2} \sim {M^3\over M^2_{pl}}
}

Not too surprisingly, it is very difficult
to come up with actual models (field theoretic or
string theoretic) which break SUSY while
simultaneously stabilizing the dilaton and
keeping the cosmological constant zero.
Despite a lot of effort very few unambiguous
results have yet been achieved.

\subsec{Dilaton Dominated SUSY breaking}

In the effective supergravity theory of a
four-dimensional superstring model, the dilaton
field can be regarded as forming a chiral
supermultiplet with the axion, dilatino, and axino:
\eqn\smult{
S = {\rm e}^{-2\phi} + ia +
\theta ({\rm dilatino}+{\rm axino})
+ \theta \theta F_S
}

Since we know that
the dilaton vev $\langle \phi \rangle$ is nonzero
it is tempting to imagine that $\langle F_S \rangle$
is also nonzero, breaking supersymmetry.

If a vev of $F_S$ is the dominant SUSY breaking
effect, we can obtain {\it model independent}
predictions for the
effective soft SUSY breaking terms
in the visible sector. This is
because, {\it in string perturbation theory},
the effective supergravity theory depends
on the dilaton superfield $S$ in a simple, universal way.
The superpotential is in fact independent of $S$
to all orders
(due to a Peccei-Quinn
symmetry of the axion), while the K\"ahler potential and
gauge kinetic functions have the following
dependence:
\eqn\kahgaug{
\eqalign{
K &= -log(S + S^*)\cr
f_a &= k_a S\cr
}}
where $k_a$ is the Kac-Moody level.

This dilaton dominated scenario has the great
virtue of making model independent predictions
of the soft SUSY breaking couplings, modulo
threshold effects:
\eqn\dilpred{
\eqalign{
m_{gauginos} &= \sqrt{3}m_{3/2}\cr
m_{scalars}^2 &= m^2_{3/2}\delta_{ij}\cr
A_{ijk} &= -\sqrt{3}m_{3/2}y_{ijk}\cr
}}
where the $y_{ijk}$ are Yukawa couplings.

Unfortunately this attractive scenario has
two important difficulties;

\noindent -- At string loop level, or if there are
significant F term vevs for other moduli, then
you get flavor-changing neutral current problems
due to nonuniversal scalar masses.

\noindent -- Nonperturbative string corrections
to (at least) the K\"ahler potential are
probably not small, which makes the
predictions suspect.

Of course, if supersymmetry is discovered and
the relations \dilpred\ appear to hold, this
will be declared a triumphant verification
of superstring theory.

\subsec{Anomalous U(1)}

In the previous section we did not consider
the possibility of D term SUSY breaking
involving the dilaton. This is in fact
an important topic, not for SUSY breaking
per se, but for other phenomenological
issues.
Many tree level conformal field theory
solutions to string theory,
including almost all known semi-realistic models,
contain a U(1) gauge
factor which is anomalous. When this occurs, the Green-Schwarz
mechanism breaks the anomalous $U(1)$, at the expense
of generating  a Fayet-Iliopoulos $D$ term proportional to
\eqn\dflat{
D_A ~=~ \sum_i Q_i^A |\chi_i|^2 + {{g^2}\over{192 {\pi}^2}} e^{\phi}
{}~ Tr Q^A ~~~~,
}
where $\phi$ is the dilaton, $TrQ^A$ is the
trace over the U(1) charges which gives the
mixed gauge-gravitational
anomaly, and the
$\chi_i$ are scalar fields with anomalous charge $Q_i^A$.
This term will break supersymmetry and destabilize
the vacuum.
The vacuum becomes stable and supersymmetry is restored when one
or more of the scalar fields which carry nonzero anomalous charge
acquire a vev such that the right-hand side of \dflat\ vanishes.
Supersymmetry is then restored provided that this vacuum shift
is in a direction which is F-flat and also D-flat with respect
to all non-anomalous $U(1)$'s. If we let $\chi_i$ now denote the
scalar {\it vevs} which cause \dflat\ to vanish, then the additional
D and F flatness constraints are
\eqn\fflat{
D_a ~= ~ \sum_i Q_i^a | \chi_i |^2 ~=~0 ,
{}~~<{{\partial W}\over{\partial \phi_j}}> ~=~0 ~~~~.
}
where $a$ labels the nonanomalous $U(1)$'s, and the $\phi_j$ are
all the chiral superfields, not just those whose scalar components
get vevs.

Note that the shifted vacuum is no longer a classical string vacuum,
but does correspond to a consistent perturbative quantum string vacuum.
Thus conformal field theory
solutions which contain an anomalous $U(1)$ in some sense
access a much larger class of perturbative string vacua than those
that do not.

Note also that because the D term cancellation in
\dflat\ involves the one-loop generated anomaly, the scale of vevs
in \dflat\ is naturally (depending on the value of the anomaly)
smaller than the string scale, by an order of magnitude or so.
Since the scalars whose vevs $\chi_i$ contribute to \dflat\
often carry a variety of other abelian and nonabelian quantum numbers,
the vacuum shift generically breaks the original gauge group
to one of smaller rank. This rank reduction is variable and can
be quite large. It may be possible to perform this vacuum shift
without breaking the standard model gauge group, although there
is no fundamental reason why this should {\it always} be the case.
In fact, in many solutions there is considerable
freedom in choosing the flat directions involved in the shift.

After the vacuum shift a number of previously massless fields
will acquire masses, of order $(\alpha_{str})^nM_{str}$
for some $n$, via coupling
to scalar vevs. The spectrum of light fields, particularly light
exotics, is often much reduced. In addition, the scalar vevs also
tend to induce a number of effective Yukawa interactions for the
MSSM quarks and leptons, with Yukawa couplings that are naturally
suppressed by powers of $\alpha_{str}$.
The fact that this combination of
favorable outcomes occurs automatically in the known
semi-realistic orbifold models and free fermionic
models is quite remarkable!

\subsec{Quark and Lepton Masses}

A major challenge for any unified model is to
reproduce, even qualitatively, the many observed hierarchies
of masses and mixings for quarks and leptons. In known
semi-realistic string models
the numerical values of the couplings in the effective superpotential
are order one, and this is likely to be true rather generally
in perturbative string vacua. Thus,
small Yukawa couplings in the MSSM may originate
from scalar vevs or fermion condensates which
take values at scales other than $M_{str}$. Nonrenormalizable
couplings of quarks and leptons to these vevs or condensates
can then generate effective Yukawa couplings which are small.
A beautiful property of the known string models
is that such a mechanism does indeed occur: the vacuum
shift associated with the anomalous $U(1)$.
Even so
it appears unlikely that any one mechanism will
explain all of the observed hierarchies.

Some semi-realistic string models
can produce a top quark Yukawa which is order one, while
all the other effective Yukawas are suppressed by an
order of magnitude or more.
This is again encouraging, since such a
feature is in not built into the
construction of these models, except for the
intriguing fact that the requirement of
precisely three generations is realized
{\it in these models} in a way which
also implies restricted and flavor-dependent
Yukawas. While one cannot take such
model dependent perturbative results very
seriously, it is tempting to speculate
that string theory may ultimately relate the heaviness
of the top quark to the number of generations!

\subsec{Fractional Charge}

All $SU(3)_c$$\times$$SU(2)_L$$\times$$U(1)_Y$
conformal field theory string solutions with
$k_3$$=$$k_2$$=1$ must contain exotics with fractional
electric charge. This is because,
if all physical states in a string vacuum obey charge quantization,
there exists a certain conformal operator which
is mutually local with respect to all the physical fields. This
operator must thus itself correspond to a physical field, which leads
to a contradiction unless the standard model gauge
group is promoted to unbroken $SU(5)$ at level one.

This argument does not determine whether or not there are
any {\it massless} string states with fractional charge;
it may be possible to arrange for fractional charges to occur
only in the {\it massive} modes of the string, and thus be superheavy.
However in all of the known
semi-realistic models, fractionally charged
exotics do occur at the massless level.
Some or all of these may become superheavy after
the vacuum shift associated with the anomalous $U(1)$;
others may be confined by nonabelian hidden sector
gauge interactions.
It may be possible to avoid fractionally charged exotics
entirely in three generation string models with higher Kac-Moody
levels, but this has never been demonstrated.

The lightest fractionally charged particle will be stable. This
can create conflicts with experimental bounds from direct searches,
as well as rather
severe cosmological and astrophysical bounds.
For example, the lightest fractionally
charged particle will completely dominate the energy density in
the universe if its mass is greater than
a few hundred Gev.
If there is an inflationary epoch and subsequent reheating, we
can probably tolerate a lightest fractionally charged particle
with mass greater than the reheating temperature.

In the known semi-realistic string models,
a variety of fractionally charged exotics
are seen to occur. They can be $SU(3)_c$$\times$$SU(2)_L$ singlets
with hypercharges less than $\pm 2$, and
they can be color triplets or Higgs
with nonstandard hypercharge. These exotics have important effects
on the RG running of the couplings.

\subsec{Rapid proton decay}

String models typically violate matter parity,
allowing for the appearance of B and L violating terms in
the cubic part of the effective superpotential. In
particular, terms of the form
\eqn\badterms{Q\,L\,d^c + u^c\,d^c\,d^c
}
where $Q$ denotes a quark doublet, $L$ a lepton doublet,
and $u^c$, $d^c$ the conjugates of the right-handed up and down quarks,
would lead to instantaneous proton decay. In addition to these cubic
terms, there is also
the possibility of quartic terms which can lead to unacceptably
rapid proton decay.

To check a particular string model for the absence of
such dangerous terms, it is insufficient to
compute the effective superpotential to quartic order.
This is because the dangerous B violating terms may
be generated at {\it any} order via nonrenormalizable terms
which are unsuppressed due to string scale vevs.
One simple solution to this problem is to
gauge $U(1)_{B-L}$;
other possibilities that have been considered in the
known models are a combination of B-L and custodial $SU(2)$
along with other flavor symmetries
which distinguish quarks from leptons.

\newsec{The Fermionic Construction
of Four-Dimensional Superstring Vacua}

{}From now on we will specialize to the properties
of four-dimensional
perturbative heterotic superstring models
obtained from the fermionic construction.
As we saw previously, this can be regarded
as a straightforward extension of the
free field conformal field theory
approach used to construct the
10-dimensional heterotic superstring.

Consider again the four-dimensional free
field construction of section 3.6. Let us
break up each of the free Weyl fermions
introduced in \newdfg\ into a pair
of free Majorana-Weyl fermions, e.g.
\eqn\wtomw{
\chi ^a(z) = \chi_1^a(z) + i \chi_2^a(z)
}
Then the complete set of world-sheet degrees
of freedom in the light cone gauge consists
of the following set of free bosons and free
Majorana-Weyl fermions:
\eqn\fulldf{
\eqalign{
&X^i(z,\bar z)\quad i=1,2\cr
&{\bar \psi}^i(\bar z)\qquad i=1,2\cr
&\lambda ^{\alpha}(\bar z)\qquad \alpha =1,2,\ldots 18\cr
&\lambda ^a(z)\qquad a=1,2,\ldots 44\cr
}}
where $i$ is a transverse spacetime vector index,
while $\alpha$ and $a$ are ``internal'' indices.

These are in fact the world-sheet degrees of freedom
for {\it any} four-dimensional
perturbative heterotic superstring model
in the fermionic construction.
As we will see shortly, such models differ
only in choices of boundary conditions (more
precisely, spin structures) and the choice of
certain phases which appear in the partition
function.

Modular invariance severely constrains these
choices in two ways:

\noindent -- A fermionic string model cannot be
modular invariant with any {\it one} choice of
boundary conditions R/NS for the $20$$+$$44$
Majorana-Weyl fermions. The Hilbert space from
which we construct the physical states must in
fact be a product of Fock spaces corresponding
to different boundary condition choices,
called spin structures, for the
various fermions. There is a nontrivial set
of rules for how to combine spin structures
to produce modular invariants.

\noindent -- Within this product space
modular invariance also puts constraints
on the physical states --GSO projections--
beyond those of world-sheet superconformal
invariance.

To understand how this works we need to consider
string theory at one-loop, i.e. on the torus.
Let us consider at one-loop the one-point function
of the identity operator, i.e. the one-loop
vacuum-to-vacuum amplitude. This has a physical
interpretation as a partition function.
By imposing modular invariance on this amplitude
we can identify and count the physical states
which survive the GSO projections.

Start with the simplest case: the partition function
of a single Majorana fermion (i.e. one left-moving
Majorana-Weyl fermion and one right-moving
Majorana-Weyl fermion. We can in fact construct
a modular invariant conformal field theory for
this case: it is not string theory but rather
describes the critical behavior of the two-dimensional
Ising model.

Now not all tori can be mapped into each other by
a conformal transformation. There is a complex
parameter, $\tau$, which parametrizes the inequivalent
tori. So one constraint of modular invariance is
that we must compute the partition function for
fixed $\tau$, then integrate in $\tau$ over some
appropriate domain:
\eqn\ztauint{
Z = \int_{{\rm Fundamental}\atop {\rm domain}}
{d^2\tau\over \tau _2}
\left[ \eta (\tau ) {\bar \eta}(\bar \tau)
\sqrt{\tau _2} \right]^{-2}
Z(\tau )
}
where $\tau$$=$$\tau _1$$+$$i\tau _2$ and $\eta (\tau )$
is the Dedekind eta function.

$Z(\tau )$ is just the path-integral of the
action over the torus parametrized by $\tau$:
\eqn\ztauis{
Z(\tau ) = \int _{\rm torus} d^2z\, {\rm e}^{-S}
= tr\;{\rm e}^{2\pi i \tau _1 P}\,
{\rm e}^{-2\pi \tau _2 H}
}
the second expression is in Hamiltonian operator
form, where
\eqn\handp{
\eqalign{
H &= L_0 + {\bar L}_0 - {1\over 24}\cr
P &= L_0 - {\bar L}_0\cr
}}
Thus we can write
\eqn\simpztau{
Z(\tau ) = q^{-1/48} {\bar q}^{-1/48}
tr\; q^{L_0} {\bar q}^{{\bar L}_0}
}
where $q$$=$exp($2\pi i\tau$).
Since we know how $L_0$ acts on states in the Fock
space, we can compute this.
Furthermore, we can compute it separately for the left and
right-moving pieces.

We have to specify two boundary conditions for
each fermion, corresponding to the two independent
cycles of the torus. If you like you may think of these
as ``space'' and ``time'', each compactified on
a circle. The ``space'' choice determines the mode
expansion of $\psi (z)$ to be either R or NS, i.e.
either periodic or antiperiodic modes. If the ``time''
boundary condition is NS, then the partition function
is just given by the naive trace with $L_0$ acting
on the appropriate R or NS Fock space:
\eqn\nsspins{
\znn (\tau ) = tr_{NS}\; q^{L_0-1/48} \quad ;
\quad \znr (\tau ) = tr_{R}\; q^{L_0-1/48}
}
When the ``time'' boundary condition is Ramond
the definition of the trace is modified:
\eqn\rspins{
\zrn (\tau ) = tr_{NS}\; (-1)^F q^{L_0-1/48} \quad ;
\quad \zrr (\tau ) = tr_{R}\; (-1)^F q^{L_0-1/48}
}
where $F$ is the fermion number operator,
defined by the relations
\eqn\fnodef{
\eqalign{
\left\{ (-1)^F,\psi _n \right\} &= 0,\quad
\forall {\rm\ modes}\cr
F\, \vert 0 \rangle _{NS} &= 0\cr
F\, \vert \uparrow \rangle _{R} &= 0\cr
F\, \vert \downarrow \rangle _{R} &= 1\cr
}}
where $\vert\uparrow\rangle$ and
$\vert\downarrow\rangle$ are the doubly
degenerate Ramond vacua.

$\znn$, $\znr$, $\zrn$, $\zrr$ define the
four possible spin structures for a single
Majorana-Weyl fermion on the torus. Modular
invariance now requires that we find
combinations of these four functions and
their right-moving analogs which are
invariant under the torus modular
transformations. These transformations are
generated by:
\eqn\tmodtran{
\eqalign{
\tau &\to \tau + 1\cr
\tau &\to -1/\tau \cr
}}
After some fiddling around with the Poisson
resummation formula, one sees that under these
transformations
\eqn\firstttran{
\eqalign{
\tau \to \tau + 1:\qquad
\znn (\tau ) &\to {\rm e}^{-{\pi i\over 24}}\zrn (\tau )\cr
\zrn (\tau ) &\to {\rm e}^{-{\pi i\over 24}}\znn (\tau )\cr
\znr (\tau ) &\to {\rm e}^{{\pi i\over 12}}\znr (\tau )\cr
\zrr (\tau ) &\to \zrr (\tau )\cr
}}
\eqn\secondttran{
\eqalign{
\tau \to -1/\tau :\qquad
\znn (\tau ) &\to \znn (\tau )\cr
\zrn (\tau ) &\to \znr (\tau )\cr
\znr (\tau ) &\to \zrn (\tau )\cr
\zrr (\tau ) &\to \zrr (\tau )\cr
}}
The right-moving analogs transform in
precisely the same way, with the
phases complex conjugated.

At first sight it appears that $\zrr (\tau )$ is
modular invariant all by itself. This is true,
however $\zrr (\tau )$ evaluates to zero, due
to the Ramond zero mode which makes the
Ramond vacuum doubly degenerate. So I can
make a modular invariant partition function this way,
but it vanishes.

So now we see that indeed we need a nontrivial
{\it combination} of spin structures to get
a modular invariant partition function. We
must combine left-movers with right-movers
and sum over different spin structures.
In our example:
\eqn\ispf{
\znn (\tau ){\bar Z}^{NS}_{NS}(\bar \tau )
+ \zrn (\tau ){\bar Z}^{R}_{NS}(\bar \tau )
+ \znr (\tau ){\bar Z}^{NS}_{R}(\bar \tau )
+ \zrr (\tau ){\bar Z}^{R}_{R}(\bar \tau )
}
is the correct modular invariant partition
function for the critical Ising model.

How do we count states in this partition
function? There are two {\it sectors}, the
Neveu-Schwarz and the Ramond. In each
sector there is a GSO projection, because
e.g. in the NS sector
\eqn\gsons{
\znn + \zrn = tr\left( 1 + (-1)^F \right) q^{L_0 - 1/48}
}
so states with odd fermion number are projected out
of the trace. Thus, if this were full-fledged string
theory, I should construct all the Virasoro physical
states and then {\it remove} those with odd fermion
number to get the true physical spectrum
from the NS sector.

We can immediately generalize these results to
handle four-dimensional heterotic superstring
models. We always have 20 right-moving Majorana-Weyl
fermions, two of which carry the transverse spacetime
index, and we always have 44 left-moving
Majorana-Weyl fermions. The left-moving part of
the fermionic contribution to the partition
function will have the form:
\eqn\zgenform{
Z(\tau ) = \sum_{{\rm spin\ structures}\atop
\{ {\vec \alpha}_i, {\vec \beta}_i \} }
C^{\alpha _i}_{\beta _i} Z^{\alpha _i}_{\beta _i}
}
where ${\vec \alpha}_i$ and ${\vec \beta}_i$
are a set of 44 component vectors denoting
spin structures. The $Z^{\alpha}_{\beta}$ are just
the corresponding product of
the single fermion spin structures $\znn (\tau )$, etc.,
while the $C^{\alpha}_{\beta}$ are phases.

Antoniadis, Bachas, and Kounnas, and independently
Kawai, Lewellen, and Tye, solved the modular
invariance constraints {\it in general} for such
models. Only a few additional constraints are
then required to ensure the full modular
invariance of the theory, not just of the
one-loop vacuum-to-vacuum amplitude.
You can rederive most of their result
yourself using \firstttran\ and \secondttran .

This general solution for which choices of
spin structures give modular invariant
conformal field theory solutions can be
expressed as some simple rules about
Ramond-Ramond overlaps of the vectors
${\vec \alpha}_i$ and ${\vec \beta}_i$,
and the choice of the phases $C^{\alpha}_{\beta}$.
We will use the following simple notation to
denote the components of
${\vec \alpha}_i$ and ${\vec \beta}_i$,
i.e., the individual boundary conditions
of the Majorana-Weyl fermions:
\eqn\notis{
\eqalign{
1&={\rm Ramond}\cr
0&={\rm Neveu-Schwarz}\cr
}}

One of the things modular invariance requires
is that, if ${\vec \alpha}_i$ and
${\vec \alpha}_j$ occur in the sum
over spin structures, then
${\vec \alpha}_i$$+$${\vec \alpha}_j$
must also occur. Here ``addition'' of
boundary condition vectors in the notation
of \notis\ is defined mod 2, i.e.
\eqn\addis{
\eqalign{
{\rm NS} + {\rm NS} &= {\rm NS}\cr
{\rm NS} + {\rm R} &= {\rm R}\cr
{\rm R} + {\rm R} &= {\rm NS}\cr
}}
It is also the case that if some
${\vec \alpha}_i$ occurs in the sum
over spin structures, then an
equivalent ${\vec \beta}$ must also
occur. These two facts taken together
imply that any allowed set of spin
structures is completely specified
by some set of {\it basis vectors}
${\vec V}_i$. The complete set of
${\vec \alpha}_i$ and ${\vec \beta}_i$
then consists of all possible distinct linear
combinations of the basis vectors with
positive integer coefficients (there are
a finite number of these since ``addition''
is mod 2).

Modular invariance also requires that
every model contains at least the
following two sectors:

\noindent -- the all-Neveu-Schwarz or ``untwisted''
sector
\eqn\untwistsec{
\left( 0^{20} \| 0^{44} \right)
}
where the double vertical line separates the
boundary conditions of the 20 right-movers
from those of the 44 left-movers.

\noindent -- the all-Ramond sector
\eqn\allram{
\left( 1^{20} \| 1^{44} \right)
}

Since the untwisted sector is obtained as twice
the all-Ramond sector, we only need the latter as
a basis vector. Conventionally this is always
labelled $V_0$:
\eqn\vzerodef{
V_0 = \left( 1^{20} \| 1^{44} \right)
}

Given a set of basis vectors consistent with
modular invariance we first have to identify
if there are any pairs of left-left Majorana-Weyl
fermions (or pairs of right-right
Majorana-Weyl fermions) whose boundary conditions
{\it match} in the entire set of
spin structures (equivalently, in all the
basis vectors). Any such pairs are actually
Weyl fermions, and we must treat them
separately since bilinears of them
are {\it currents} (recall the 16 Weyl pairs
in the 10-dimensional heterotic string
which generated $SO(32)$).

We should then identify any right-left pairs
whose boundary conditons match in all the
basis vectors. These pairs make copies
of the Ising model discussed aboved; we thus call
them Ising fermions.

Whatever is left are by definition
Majorana-Weyl fermions whose
boundary conditions {\it don't} pair
up with any other
Majorana-Weyl fermion.
These unpaired Majorana-Weyl fermions
have been called ``real fermions''
or ``chiral Ising'' fermions in the
literature.

A partition function will be a sum over
sectors, labelled by ${\vec \beta}_i$.
The contribution of each sector is itself
a sum over ${\vec \alpha}_i$, which has
the effect of performing the GSO projections.
The number of independent GSO projections
performed for each sector is
approximately equal to the number
of basis vectors. For semi-realistic models
which have hundreds of sectors and perhaps
ten basis vectors, you obviously need a computer
to extract the physical spectrum.

\subsec{$SO(32)$ Versus $E_8$$\times$$E_8$ in 10 Dimensions}

We now know enough formalism to understand the
difference between the $SO(32)$ and
$E_8$$\times$$E_8$ heterotic strings in ten dimensions.
The modular invariance rules for ten dimensions
are the same as in four dimensions. The only
difference in the construction is the counting
of Majorana-Weyl fermions: there are now only
8 right-movers, all of which carry a transverse
spacetime index, and there are only 32 left-movers.

The $SO(32)$ heterotic string can be defined
(there are many equivalent definitions) by the
following two basis vectors:
\eqn\sobv{
\eqalign{
V_0 &= \left( 1^{8} \| 1^{32} \right) \cr
V_1 &= \left( 1^{8} \| 0^{32} \right) \cr
}}
Let us locate all the massless physical states.
The untwisted sector contains the graviton,
dilaton, antisymmetric tensor, and the 496
gauge bosons of $SO(32)$, obtained
from currents which are bilinears
of the 16 left-moving Weyl fermions.
This gauge group is realized at Kac-Moody
level one, as is always the case when
the currents are all fermion bilinears.
Sector $V_1$ contains the gravitino,
dilatino, tensorino, and the gauginos
--recall from section 3.3 that to
get {\it spacetime} fermions we need
a spin field, and thus we need
a sector which is Ramond in the right-moving
slots whose associated fermion carries
a spacetime index. There are two other
sectors in this model, $V_0$ and $V_0$$+$$V_1$,
but neither of these contains any massless
physical states.

The $E_8$$\times$$E_8$ heterotic string can
be defined by the following three basis
vectors:
\eqn\eightbv{
\eqalign{
V_0 &= \left( 1^{8} \| 1^{32} \right) \cr
V_1 &= \left( 1^{8} \| 0^{32} \right) \cr
V_2 &= \left( 0^{8} \| 1^{16} 0^{16} \right) \cr
}}
Let us locate all the massless physical states.
The untwisted sector contains the graviton,
dilaton, antisymmetric tensor, and the 240
gauge bosons of $SO(16)$$\times$$SO(16)$, obtained
from currents which are bilinears of either
the first 8 left-moving Weyl fermions, or of
the second 8 left-moving Weyl fermions.
Sector $V_2$ contains another 128 gauge bosons,
obtained from currents which are composites of
left-moving {\it twist fields}, not fermion
bilinears.
To be precise, there is a pair of
twist fields $\sigma ^{\alpha}(z)$, $\alpha$$=$$1$, $2$
for each of the 8 left-moving Weyl fermions
in $V_2$ that have Ramond boundary conditions.
The vertex operators of the gauge bosons coming
from $V_2$ have the form
\eqn\vovtwo{
{\bar \psi}^{\mu}(\bar z)
\sigma ^{\alpha _1}_1 \sigma ^{\alpha _2}_2
\sigma ^{\alpha _3}_3 \sigma ^{\alpha _4}_4
\sigma ^{\alpha _5}_5 \sigma ^{\alpha _6}_6
\sigma ^{\alpha _7}_7 \sigma ^{\alpha _8}_8
{\rm e}^{ip\cdot X}
}
There are $2^8$ such states (not counting helicities)
before the GSO projections.
$V_1$ provides no projection on these states, since
$V_1$ and $V_2$ have no Ramond-Ramond overlap.
$V_0$ and $V_2$ provide the {\it same} GSO
projection, which removes half the states. Thus
we get $2^7$$=$$128$ gauge bosons from sector $V_2$.
Sector $V_0$$+$$V_1$$+$$V_2$ also
gives another 128 gauge bosons, filling out
the $240$$+$$128$$+$$128$$=$$496$ gauge bosons
of $E_8$$\times$$E_8$, at Kac-Moody level one.
The corresponding gauginos
come from sectors $V_1$, $V_1$$+$$V_2$, and
$V_0$$+$$V_2$.

Note that the $E_8$$\times$$E_8$ heterotic string
was obtained by simply adding one additional
basis vector, $V_2$, to the set that gave the
$SO(32)$ heterotic string. This additional basis
vector had two distinct effects on the massless
physical spectrum. The first effect was to
provide an additional GSO projection on the
untwisted sector and sector $V_1$, which
removed the 248 gauge bosons and gauginos
of $SO(32)$ not in
$SO(16)$$\times$$SO(16)$. The second
effect was to provide new massless states,
in sector $V_2$ itself and in sector
$V_0$$+$$V_1$$+$$V_2$.

This is the general pattern for model-building
in the fermionic construction in any number
of dimensions. Each additional basis vector
(usually) provides both new massless states
and new GSO projections which eliminate some
of the previous set of massless states.
Because these two effects are highly
correlated, and constrained by the
modular invariance rules,
it becomes something of an art
to construct sets of basis vectors that
will produce a desired massless spectrum
and gauge group.

\newsec{Examples of Semi-Realistic Models}

\subsec{The Flipped $SU(5)$ Model}

The flipped $SU(5)$ model was the first
semi-realistic model to be obtained in the
fermionic construction. It has a number of
simplifying features, e.g. it contains only
Weyl and Ising fermions, and uses a very simple
gauge embedding of $SU(5)$.

There a several slightly different versions
of the flipped $SU(5)$ model; we will discuss
the ``search'' version, so-called because it
appears in a paper with that word in the title.
While it is a priori unlikely that this model
corresponds to the real world, it is not
obviously wrong. As discussed already in
section 4, a number of happy circumstances
seem to conspire to make the model look
much more like the real world than one
would have expected from the weak criteria
of ``semi-realistic''.

The model is defined by 8 basis vectors --
I will suppress in this discussion the
additional phases $C^{V_i}_{V_j}$
which must also be specified.
It will be more illuminating to first discuss
the model defined by just the first 5 of these
basis vectors:
\eqn\fivebv{
\eqalign{
V_0 &= \left( (1^2)(111111)
\hbox to38pt{\hfil$(1^{12})$\hfil}\,\|\,
(1^{10})(111111)
\hbox to38pt{\hfil$(1^{12})$\hfil}(1^{16}) \right) \cr
V_1 &= \left( (1^2)(111111)
\hbox to38pt{\hfil$(0^{12})$\hfil}\,\|\,
(0^{10})(000000)
\hbox to38pt{\hfil$(0^{12})$\hfil}(0^{16}) \right) \cr
V_2 &= \left( (1^2)(110000)
\hbox to38pt{\hfil$(1^{4}0^{8})$\hfil}\,\|\,
(1^{10})(110000)
\hbox to38pt{\hfil$(1^{4}0^{8})$\hfil}(0^{16}) \right) \cr
V_3 &= \left( (1^2)(001100)
\hbox to38pt{\hfil$(0^{4}1^{4}0^{4})$\hfil}\,\|\,
(1^{10})(001100)
\hbox to38pt{\hfil$(0^{4}1^{4}0^{4})$\hfil}(0^{16}) \right) \cr
V_4 &= \left( (1^2)(000011)
\hbox to38pt{\hfil$(0^{8}1^{4})$\hfil}\,\|\,
(1^{10})(000011)
\hbox to38pt{\hfil$(0^{8}1^{4})$\hfil}(0^{16}) \right) \cr
}}

Note that all of the Majorana-Weyl fermions
in this model pair up into Weyl fermions.
The model defined by just the first two of these
basis vectors is the 10-dimensional $SO(32)$ heterotic
string compactified on a six-torus, with each circle
of the torus fixed at the fermionic radius. The gauge
group of such a solution is $SO(32)$$\times$$[U(1)]^{12}$
at generic values of the radii, but at the fermionic radii
it is enhanced to $SO(44)$$\times$$[U(1)]^6$,
with Kac-Moody level one.
The model defined by just the first two
basis vectors also has $N$$=$$4$ spacetime
supersymmetry, i.e. there are 4 gravitinos in
the massless spectrum.

Each additional basis vector $V_2$, $V_3$, or $V_4$
adds GSO projections which remove half of the gravitinos;
however the projection from $V_4$ is equivalent to
the projection from $V_0$$+$$V_2$$+V_3$, and thus one
gravitino survives. This means that the model defined
by \fivebv\ has $N$$=$$1$ spacetime supersymmetry.
The resulting model can be thought of as a symmetric
orbifold of the torus model.
The 22 left-moving Weyl fermions no longer contribute
the currents of $SO(44)$ realized as fermion bilinears
in the untwisted sector.
Rather, from \fivebv\ it is
clear that only the subgroup
$SO(10)$$\times$$[SO(6)]^3$$\times$$SO(16)$
arises from bilinears in the untwisted sector.
In addition, there are $2^8$ new currents
coming from twist fields in sector
$V_0$$+$$V_2$$+$$V_3$$+$$V_4$; there is
only one distinct GSO projection on these,
leaving 128 additional gauge bosons. This
promotes the $SO(16)$ to an $E_8$, still at
Kac-Moody level one.

Once we have identified the full gauge group, and
precisely how this group is embedded into quantum
numbers of the fermionic Fock space, we can
determine how all of the massless states
transform under the group. We are guaranteed
(by unitarity or modular invariance)
that all of the massless states assemble
into a set of complete irreducible representations
of the gauge group.

In addition to the gauge bosons, the gravity multiplet,
and their superpartners, we get massless fermions
from sectors $V_2$, $V_3$, and $V_4$ (their scalar
superpartners come from $V_1$$+$$V_2$, $V_1$$+$$V_3$,
and $V_1$$+$$V_4$). Let us count the number of massless
fermion states coming from, e.g. sector $V_2$. The
massless states correspond to all of the Weyl fermions
being in the vacuum state; for each Ramond Weyl fermion
there are two degenerate vacuum states, as we have already
discussed. The first two right-mover slots of $V_2$
correspond to the two Majorana-Weyl fermions which
carry the transverse spacetime index; the fact that
these are Ramond for $V_2$ tells us that the
massless states are spacetime fermions, with two
possible helicities. The total number of massless
states from $V_2$ before the GSO projections is
2 helicities times a $2^3$ right-mover Ramond degeneracy
times a $2^5$ degeneracy in the left-mover $SO(10)$ slots
times a $2^3$ degeneracy in the left-mover $SO(6)$ slots.
For fixed helicity, these states correspond to
8 copies of a
$(16,4)$$+$$(\bar{16},4)$$+$$(16,\bar 4)$$+$$(\bar{16},\bar 4)$
of $SO(10)$$\times$$SO(6)$.
The GSO projection from $V_3$ makes these fermions
{\it chiral} with respect to $SO(10)$.
This is clear from \fivebv , where
we see that the Ramond-Ramond overlaps of $V_2$ with
$V_3$ occur only in the first two right-movers and
the $SO(10)$ slots of the left-movers.
After this projection the {\it positive helicity}
states consist of 8 copies of a
$(16,4)$$+$$(16,\bar 4)$; their CPT conjugates
are negative helicity states in a
$(\bar{16},4)$$+$$(\bar{16},\bar 4)$.
Continuing, the GSO projections from $V_1$ and from
$V_2$ itself reduce the 8 copies down to 2 copies;
$V_0$ and $V_4$ do not provide any new projections.
Thus we find that after all the projections
$V_2$ contibutes 2 copies of chiral fermions
in a $(16,4)$$+$$(16,\bar 4)$ of
$SO(10)$$\times$$SO(6)$.

If we think of this $SO(6)$
as a gauged flavor symmetry, and recall that a
chiral 16 of $SO(10)$ contains precisely one
generation of standard model fermions plus a right-handed
neutrino, then we may say that $V_2$ contributes
16 light generations. $V_3$ and $V_4$ similarly
contribute 16 generations each, for a total
of 48 generations.

This is very encouraging, since we want 3 generations
for a semi-realistic model. Clearly all we need now
is to introduce 4 additional GSO projections on
$V_2$, $V_3$, and $V_4$, to cut the number of states
down to one generation each.

We must also do something about the gauge group.
Without worrying yet about the ``flavor'' group
$[SO(6)]^3$ or the ``hidden'' group $E_8$, there
is a problem with leaving $SO(10)$ unbroken.
The problem is that there are no massless adjoint
Higgs chiral supermultiplets in the spectrum
to break $SO(10)$ to
$SU(3)_c$$\times$$SU(2)_L$$\times$$U(1)_Y$
as in a standard field theoretic GUT.

Let us examine why this is. We know that we have
a set of left-moving currents in the adjoint
representation of $SO(10)$, formed from Weyl
fermion bilinears. So why can't we make massless
adjoint Higgs from these? The vertex operator
for such adjoint Higgs has the form
\eqn\adjvo{
\lambda ^{\alpha}(\bar z)\lambda ^a(z)
\lambda ^b(z) {\rm e}^{ip\cdot X}
}
Here $a$ and $b$ label the bilinears that
make the $SO(10)$ currents, while
$\alpha$$=$$1,2,\ldots 9$ labels any one of
the ``internal'' right-moving Weyl fermions.
These vertex operators will indeed create
adjoint Higgs in untwisted sector,
before the GSO projections. The GSO projection
from $V_1$, however, only allows the states
where $\alpha$ labels one of the 3
right-mover Weyl fermions which are Ramond
in $V_1$; let us call these
$\alpha$$=$$1,2,3$.
Furthermore, this model has
$N$$=$$1$ spacetime SUSY, not
$N$$=$$4$.
This implies that there
is at least one GSO projection
in the untwisted sector which
distinguishes the spacetime helicity
slots from each
$\alpha$$=$$1,2,3$.
These projections {\it either} remove
$SO(10)$ gauge bosons, or $SO(10)$
adjoint Higgs. Since by assumption
they do not remove any $SO(10)$
gauge bosons, all of the $SO(10)$
adjoint Higgs are projected out.

We have proven that we cannot make adjoint
Higgs from currents, but perhaps there are
other left-moving local
conformal operators in the adjoint
representation of $SO(10)$ that we
could use to make massless adjoint
Higgs. Such operators, to make massless
physical states, have to be what are called
{\it primary} conformal operators with
respect to the Kac-Moody algebra. However
for any Kac-Moody algebra at level one
there are no primary conformal operators
in the adjoint.

More generally, for $SO(10)$ at level one
the only massless chiral supermultiplets
which can occur are the singlet, the 10,
the 16, and the $\bar{16}$. For $SO(10)$
at level two we can have in addition
the 45 and the 54. Similarly, for
$SU(5)$ at level one we can only
have the singlet, the 5, $\bar 5$, 10,
and $\bar{10}$. For $SU(5)$ at level two
we can have in addition the 24, 40,
$\bar{40}$, 45, and $\bar{45}$.

This long detour has taught us that,
at the same time as we cut the number of
generations in \fivebv\ from 48 to 3,
we must break $SO(10)$ to either
$SU(3)_c$$\times$$SU(2)_L$$\times$$U(1)_Y$
or to flipped $SU(5)$$\times$$U(1)$.
Flipped $SU(5)$ avoids the necessity of
adjoint Higgs by embedding $U(1)_Y$
in $SU(5)$$\times$$U(1)$ differently than
the standard $SU(5)$ embedding; this has the
result that the 10 and $\bar{10}$ of
$SU(5)$ can now get vevs which are standard model
singlets, and accomplish the desired
symmetry breaking.

Now we can write down the final three basis
vectors which, together with \fivebv ,
define the flipped $SU(5)$ model:
\eqn\fivebv{
\eqalign{
V_5 &= \left( (1^2)(110000)
\hbox to56pt{\hfil$(1^20^210^310^3)$\hfil}\,\|\,
(1^{10})(110000)
\hbox to56pt{\hfil$(1^20^210^310^3)$\hfil}(0^{16}) \right) \cr
V_6 &= \left( (1^2)(001100)
\hbox to56pt{\hfil$(10^31^20^310^2)$\hfil}\,\|\,
(1^{10})(001100)
\hbox to56pt{\hfil$(10^31^20^310^2)$\hfil}(0^{16}) \right) \cr
V_7 &= \left( (1^2)(000000)
\hbox to56pt{\hfil$(10^510^5)$\hfil}\,\|\,
(\ha ^{10})(\ha\ha\ha\ha\ha\ha)
\hbox to56pt{\hfil$(10^510^5)$\hfil}(\ha ^81^40^4) \right) \cr
}}
It is evident that
not all of the Majorana-Weyl fermions pair up
into Weyl fermions -- in fact there are now
8 Ising fermions.
The notation \ha\ in $V_7$ is defined as follows.
As we pointed out previously, each Weyl fermion
may be regarded as a complexified pair of
Majorana-Weyl fermions:
\eqn\wtomwb{
\lambda (z) = \lambda _1(z) + i \lambda _2(z)
}
We therefore may choose from a more general
set of boundary conditions for Weyl fermions,
corresponding to rotating $\lambda _1$, $\lambda _2$
into each other:
\eqn\moregenbc{
\left( {\lambda _1\atop \lambda _2} \right)
\to \left( \matrix{cos\pi\theta &sin\pi\theta\cr
		-sin\pi\theta &cos\pi\theta\cr}
\right) \left( {\lambda _1\atop \lambda _2} \right)
}
In this language $\theta$$=$$0$, $1$ correspond to
NS, R boundary conditions, while
$\theta$$=$$1/2$ is the new boundary condition
we have used in $V_7$.

The GSO projections from $V_5$, $V_6$, and $V_7$ cut the
number of generations from 48 to 24 to 12 to 6. In
addition, there is a independent GSO projection
associated with $2$$*$$V_7$,
\eqn\twotsev{
2*V_7 = \left( 0^{20} \,\|\, 1^{16}0^{12}1^{8}0^{8}
\right)
}
which cuts the number of generations down to three,
one from each of $V_2$, $V_3$, and $V_4$.
Of course since we have introduced new sectors
we must also worry that new generations appear
in the new sectors. Indeed in this model there
is new $SU(5)$ matter in the sectors, but it is
all vectorlike; thus the net number of generations
is still three.

The full gauge group is
\eqn\fullgg{
\left[ SU(5)\times U(1) \right]_{\rm flipped}
\times \left[ U(1) \right]^5_{\rm flavor}
\times \left[ SO(10) \times SO(6) \right]_{\rm hidden}
}
where ``flavor'' indicates that the standard model
fermions carry nonzero flavor dependent charges
under these five extra $U(1)$s. One combination
of these $U(1)$s is anomalous; we must therefore
perform a vacuum shift as described in section 4.5.
The analysis of all the F and D flat directions is
too complicated to go into here. The vacuum shift
will in general break some
or all of the extra $U(1)$s at a scale of
about $10^{16}$ GeV.

Here is a complete list of the massless chiral
superfields in this model:

\noindent -- five 10's of $SU(5)$ (with various
extra $U(1)$ charges).

\noindent -- two $\bar{10}$'s which combine with
two combinations
of the five $10$'s above to make two vectorlike pairs.
One of these pairs is presumably the Higgs which break
flipped $SU(5)$ down to
$SU(3)_c$$\times$$SU(2)_L$$\times$$U(1)_Y$,
although the perturbative superpotential
of the effective field theory below the string
scale gives no indication of this.
The other pair may be superheavy after the vacuum shift.

\noindent -- three $\bar 5$'s.

\noindent -- four vectorlike pairs of
$5$$+$$\bar 5$'s. One linear combination of
these can be the MSSM Higgs; the others are
exotics.

\noindent -- 27 $SU(5)$$\times$$SO(10)$$\times$$SO(6)$
singlets carrying various $U(1)$ charges.

\noindent -- three 10's of the hidden $SO(10)$.

\noindent -- seven 6's and six $4$$+$$\bar 4$ pairs
of the hidden $SO(6)$.

By computing n-point functions in this
conformal field theory of Weyl and Ising fermions
we can determine the effective field theory
below the string scale. The effective
superpotential has no linear or quadratic
terms (because of conformal invariance), 53 cubic
terms, 15 quadratic terms, and over 1000 quintic
terms. The proliferation of terms at quintic
and higher order is a generic feature of
semi-realistic models; fortunately only a limited
number of terms at any order represent new
physically important contact interactions -- the
rest are either physically uninteresting or
represent diagrams which are composites of many
lower order vertices. However to do a complete
analysis of F and D flat directions for such
an effective superpotential is still a
daunting task even with a computer.

This model has Yukawa terms in the cubic
superpotential, as well as effective Yukawas
in the quartic and higher terms which arise
when various $SU(5)$ singlet fields get vevs
in the vacuum shift. The cubic Yukawa couplings
are order one, while the higher order effective
Yukawas are suppressed by powers of a parameter
which is approximately $1/10$. Without committing
ourselves to a specific vacuum shift there is
not much more that we can say in detail.
However one can immediately read off from the
cubic superpotential that at most one of the
3 up-type quarks (u,c,t), and at most one
of the down-type quarks (d,s,b),
and at most one of the charged
leptons (e,$\mu$,$\tau$) has an unsuppressed
Yukawa coupling. Thus this model at the very
least has a natural hierarchy between the
third generation and the two light generations.

Clearly a lot more detailed analysis is
needed to make a hard comparision between
this model and the real world. This has been
accomplished to a large extent for two other
classes of semi-realistic models: the $Z_3$
orbifold models, and the
$SU(3)_c$$\times$$SU(2)_L$$\times$$U(1)_Y$
fermionic models of Faraggi
-- which are based on the same construction
as flipped $SU(5)$. Flipped $SU(5)$ has
an additional ambiguity relative to these
models, because the details of
the $SU(5)$ breaking are not determined in
string perturbation theory. However all
three classes of models are otherwise
very similar in character. They all have
potential difficulties with gauge coupling
unification, fractionally charged particles,
masses and mixings of the lighter generations,
and providing an appropriate SUSY breaking
mechanism. But as a first pass at realistic
superstring phenomenology, they succeed
remarkably well.

\subsec{A Three Generation $SU(5)$ Model}

We saw in the previous subsection that you
cannot have a semi-realistic string model
which resembles a conventional
$SU(5)$ GUT unless the $SU(5)$ current
algebra is realized at Kac-Moody level
greater than one. In the fermionic
construction it is only possible to
achieve Kac-Moody levels 1, 2, 4, and 8.
Going to level two or larger allows adjoint
Higgs in the massless spectrum, but also
allows the possibility of new exotics in
higher dimensional irreps of the GUT group.

We will begin with a construction which
produces $SO(16)$ at Kac-Moody level two.
This can be done with the following
six basis vectors:
\eqn\sixbv{
\eqalign{
V_0 &= \left( (1^2)(111111)
\hbox to27pt{\hfil$(1^{12})$\hfil}\,\|\,
(1111111111111111)(1111111111111111)
(0^{12}) \right) \cr
V_1 &= \left( (1^2)(111111)
\hbox to27pt{\hfil$(0^{12})$\hfil}\,\|\,
(0000000000000000)(0000000000000000)
(0^{12}) \right) \cr
V_2 &= \left( (0^2)(000000)
\hbox to27pt{\hfil$(0^{12})$\hfil}\,\|\,
(1111111100000000)(1111111100000000)
(0^{12}) \right) \cr
V_3 &= \left( (0^2)(000000)
\hbox to27pt{\hfil$(0^{12})$\hfil}\,\|\,
(1111000011110000)(1111000011110000)
(0^{12}) \right) \cr
V_4 &= \left( (0^2)(000000)
\hbox to27pt{\hfil$(0^{12})$\hfil}\,\|\,
(1100110011001100)(1100110011001100)
(0^{12}) \right) \cr
V_5 &= \left( (1^2)(110000)
\hbox to27pt{\hfil$(1^40^8)$\hfil}\,\|\,
(\pha\pha\pha\pha\mha\mha\mha\mha
\mha\mha\mha\mha\pha\pha\pha\pha)(101010101010101010)
(0^{12}) \right) \cr
}}
where the notation ``$+$'' and ``$-$'' in
$V_5$ denotes $+$$1/2$ and $-$$1/2$.

Currents from fermion bilinears always give
a level one Kac-Moody algebra, unless they
are abelian currents. Thus to make $SO(16)$
at level two only the 8 Cartan currents,
at most, can come from fermion bilinears in
the untwisted sector. In \sixbv\ we have
basically started with two level one $SO(16)$s
coming from fermion bilinears in the untwisted
sector, then introduced projections to remove
all but the $8$$+$$8$ Cartan currents. The final
basis vector, $V_5$, then removes the second 8
of these Cartan currents. Simultaneously, the
new sectors $V_2$, $V_3$, $V_4$, and
$2$$*$$V_5$ themselves contain new currents
constructed entirely from twist fields. The end
result is a single $SO(16)$ at level two.
There is also an $SO(13)$ at level one
associated with the last 13 of the 44 left-movers.

Note that in this model the first 16 left-movers
make 8 Weyl fermions, and the last 12 left-movers
make another 6 Weyl fermions. However the remaining
16 left-movers are {\it unpaired Majorana-Weyl fermions}.
The presence of unpaired Majorana-Weyl fermions is
necessary (though not sufficient) in the fermionic
construction to realize groups at higher level.

Now we add four more basis sectors to \sixbv .
These break the level two
$SO(16)$, which was embedded in the first
8 left-moving Weyl fermions, down to its
maximal subgroup $SU(8)$$\times$$U(1)$.
$V_9$ then breaks this level two
$SU(8)$ down to its maximal subgroup
$SU(5)$$\times$$SU(3)$$\times$$U(1)$.
Both the $SU(5)$ and the horizontal flavor
group $SU(3)$ are realized at level two.

\eqn\lastbv{
\eqalign{
V_6 &= \left( (0^2)(111100)
\hbox to44pt{\hfil$(0^41^40^4)$\hfil}\,\|\,
(1111111111111111)(0000000000000000)
\hbox to25pt{\hfil$(0^{12})$\hfil} \right) \cr
V_7 &= \left( (0^2)(001111)
\hbox to44pt{\hfil$(1^20^21^20^6)$\hfil}\,\|\,
(0000000000000000)(0000000000000000)
\hbox to25pt{\hfil$(1^80^4)$\hfil} \right) \cr
V_8 &= \left( (0^2)(000000)
\hbox to44pt{\hfil$(0^{12})$\hfil}\,\|\,
(1111000011110000)(0000111100001111)
\hbox to25pt{\hfil$(0^41^8)$\hfil} \right) \cr
V_9 &= \left( (0^2)(110011)
\hbox to44pt{\hfil$(1^20^41^20^4)$\hfil}\,\|\,
(1111000011110000)(1001011010010110)
\hbox to25pt{\hfil$(0^{12})$\hfil} \right) \cr
}}

For completeness, here is the matrix of $k_{ij}$s
which, in the notation and conventions of
Kawai, Lewellen, Schwartz, and Tye, determines
the phases in the partition function:
\eqn\kijmat{\left(\matrix{
\hfil 0\hfil &\hfil 0\hfil &\hfil
0\hfil &\hfil 0\hfil &\hfil 0
\hfil &\hfil 0\hfil &\hfil 0\hfil &\hfil 0\hfil &\hfil
0\hfil &\hfil 0\cr
\hfil 0\hfil &\hfil 0\hfil &\hfil 0\hfil &\hfil 0\hfil &\hfil
0\hfil & \hfil -1/2\hfil & \hfil -1/2\hfil & \hfil -1/2
\hfil &\hfil 0\hfil &\hfil 0\cr
\hfil 0\hfil &\hfil 0\hfil &\hfil 0\hfil &\hfil 0\hfil &\hfil
0\hfil & \hfil -1/2\hfil &\hfil 0\hfil &\hfil 0\hfil &\hfil
0\hfil &\hfil 0\cr
\hfil 0\hfil &\hfil 0\hfil &\hfil 0\hfil &\hfil 0\hfil &\hfil
0\hfil & \hfil -1/2\hfil &\hfil 0\hfil &\hfil 0\hfil &\hfil
0\hfil & \hfil -1/2\cr
\hfil 0\hfil &\hfil 0\hfil &\hfil 0\hfil &\hfil 0\hfil &\hfil
0\hfil & \hfil -1/2\hfil &\hfil 0\hfil &\hfil 0\hfil &\hfil
0\hfil &\hfil 0\cr
\hfil 0\hfil &\hfil 0\hfil &\hfil 0\hfil &\hfil 0\hfil &\hfil
0\hfil & \hfil -1/4\hfil & \hfil -1/2\hfil & \hfil -1/2
\hfil & \hfil -1/2\hfil &\hfil 0\cr
\hfil 0\hfil &\hfil 0\hfil &\hfil 0\hfil &\hfil 0\hfil &\hfil
0\hfil & \hfil +1/4\hfil & \hfil -1/2\hfil & \hfil -1/2
\hfil &\hfil 0\hfil & \hfil -1/2\cr
\hfil 0\hfil &\hfil 0\hfil &\hfil 0\hfil &\hfil 0\hfil &\hfil
0\hfil & \hfil +1/4\hfil &\hfil 0\hfil &\hfil 0
\hfil & \hfil -1/2\hfil & \hfil -1/2\cr
\hfil 0\hfil &\hfil 0\hfil &\hfil 0\hfil &\hfil 0\hfil &\hfil
0\hfil &\hfil 0\hfil &\hfil 0\hfil &\hfil
0\hfil & \hfil -1/2\hfil &\hfil 0\cr
\hfil 0\hfil & \hfil -1/2\hfil &\hfil
0\hfil &\hfil 0\hfil &\hfil 0
\hfil &\hfil 0\hfil &\hfil 0\hfil &\hfil 0
\hfil & \hfil -1/2\hfil & \hfil -1/2\cr
}\right)
}

This model has $N$$=$$1$ spacetime supersymmetry.
The full gauge group is
\eqn\myfullgg{
SU(5)\times
\left[ SU(3) \times
[U(1)]^2 \right]_{\rm flavor}
\times \left[ SO(5) \times [SU(2)]^4 \right]_{\rm hidden}
}
There is no anomalous $U(1)$ in this model.

Here is a complete list of the massless chiral
superfields in this model:

\noindent -- an adjoint 24 of $SU(5)$ which has zero
charge under the extra $U(1)$s.

\noindent -- three 10's of $SU(5)$ which are
in a triplet of the horizontal $SU(3)$, and
which are charged under the extra $U(1)$s.

\noindent -- three $\bar 5$'s of $SU(5)$
which are in an antitriplet of the
horizontal $SU(3)$, and
which are charged under the extra $U(1)$s.

\noindent -- two vectorlike pairs of
$10$$+$$\bar{10}$ of $SU(5)$.

\noindent -- one vectorlike pair of
$15$$+$$\bar{15}$ of $SU(5)$.

\noindent -- an $SU(5)$ anomaly-free
collection of
{\it chiral} exotics: one $\bar{15}$ of
$SU(5)$, three 5's which form an $SU(3)$
triplet, three 5's which form an $SU(3)$
antitriplet, four 5's which form two
doublets under one of the $SU(2)$s,
and one $\bar 5$.

\noindent -- a collection of $SU(5)$
singlets which transform under the
horizontal $SU(3)$ (and cancel the $SU(3)$
anomalies when combined with the above):
an adjoint 8, a vectorlike $6$$+$$\bar 6$ pair,
a chiral $\bar 6$, four triplets which form
two doublets under another of the $SU(2)$s,
a vectorlike $3$$+$$\bar 3$ pair, and two
chiral antitriplets.

\noindent -- two spinors and a vector under
the hidden $SO(5)$, five doublet-doublets
under various pairs of $SU(2)$s, one singlet
which carries only a $U(1)$ charge, and
one neutral singlet.

This model demonstrates a number of important
facts. First of all, it shows that it is possible
to obtain a three generation $SU(5)$ GUT with
adjoint Higgs from a perturbative superstring vacuum.
Secondly, it shows that it is possible in string
theory to obtain a nonabelian horizontal flavor
symmetry, with the three generations in
three-dimensional irreps of this symmetry.
This extra $SU(3)$ symmetry is optional; one
can construct a three generation $SU(5)$ model
that has only $U(1)$ gauged flavor symmetries.

On the negative side,
this model shows that going to Kac-Moody level
two can indeed result in the presence of
dangerous exotics in larger irreps,
in this case a chiral
$\bar{15}$ (plus nine associated 5's to cancel
its $SU(5)$ anomaly). This makes the
present model unrealistic; a more important
question, however, is whether obtaining
three generations in a string GUT {\it implies}
chiral exotics. We can examine this question
by looking at the underlying $SU(8)$ structure.
Three generations requires an odd number of
10's of $SU(5)$, which in turn requires an odd
number of 28's of $SU(8)$. However, {\it for
the particular fermionic embedding of} $SU(8)$
{\it used in this model}, we automatically make
a massless $\bar{36}$ of $SU(8)$ every time we make a
28. Because the $\bar{36}$ contains a $\bar{15}$
of $SU(5)$, we end up with the unwanted exotic.

So we see that in this model there is a relation
between the number of generations and the
presence of certain exotics; {\it however}
this relation appears to be an artifact of
(1) embedding $SU(5)$ in $SU(8)$, and (2)
embedding the root lattice of $SU(8)$ into
fermionic charges in a particular way.
Thus we strongly suspect that this problem
is not generic.

Another problem of this model can be seen
by computing the terms in the superpotential
involving just the $SU(5)$ adjoint scalars, or
terms involving adjoint scalars and $SU(5)$
singlets (which may get vevs). These are the
terms which generate the effective self-couplings
of the adjoint scalars from the superpotential.
The surprising result is that, at least through
quintic order, there are no such terms. In other
words, the adjoint scalars have a flat potential
-- they are moduli. Actually this property had been
noticed before in orbifold versions of
(even generation) stringy GUTs, and a similar
problem occurred in flipped $SU(5)$,
so perhaps this problem {\it is} generic.

Let me conclude with the following observation.
I have presented two GUT-like models here
only because their construction is easier to
describe and their massless spectra are simpler.
The more generic class of semi-realistic string
models break to
$SU(3)_c$$\times$$SU(2)_L$$\times$$U(1)_Y$
at the string scale. An interesting
open question in superstring
phenomenology is whether there exists
any perturbative string solution whose
spectrum and gauge group just below the string scale
is exactly that of the MSSM plus hidden
fields. The easiest way to answer this
question in the affirmative is
to construct a model.

\vskip .3in
\centerline{\bf Acknowledgements}

I would like to thank the ICTP, the organizers
of the Summer School, and particularly
Qaisar Shafi, for their hospitality.

\vfil\eject
\leftline{\bf An Incomplete Bibliography}
\vskip .5in
\noindent{\it General String Perturbation Theory
and Conformal Field Theory:}
\vskip .3in

\noindent --M. Green, J. Schwarz, E. Witten,
{\it Superstring Theory},
Cambridge press.

\noindent --J. Polchinski, ``What is string theory?'',
1994 Les Houches Summer School lectures, hepth/9411028.

\noindent --A. Belavin, A. Polyakov,
A. Zamolodchikov, Nucl. Phys. B241 (1984) 333.

\noindent --P. Goddard, D. Olive,
Int. J. Mod. Phys. A1 (1986) 303.

\noindent --P. Ginsparg,
Phys. Lett. B197 (1987) 139;
``Applied Conformal Theory'',
1988 Les Houches lectures.

\noindent --D. Gross and J. Sloan,
Nucl. Phys. B291 (1987) 41.

\noindent --V. Kaplunovsky,
Nucl. Phys. B307 (1988) 145; erratum
Nucl. Phys. B382 (1992) 436.

\noindent --M. Dine, N. Seiberg, E. Witten,
Nucl. Phys. B289 (1987) 589.

\noindent --D. Lust, S. Theisen, G. Zoupanos,
Nucl. Phys. B296 (1988) 800.

\noindent --A. Schellekens (ed),
{\it Superstring Construction},
North Holland press; Phys. Lett. B237
(1990) 363.

\noindent --M. Dine, ``Topics in string phenomenology'',
Talk at Strings93, hepph/9309319.

\noindent --T. Banks and M. Dine,
Phys. Rev. D50 (1994) 7454.

\vskip .3in
\noindent{\it The Fermionic Construction:}
\vskip .3in

\noindent --H. Kawai, D. Lewellen, H. Tye,
Nucl. Phys. B288 (1987) 1.

\noindent --I. Antoniadis, C. Bachas, C. Kounnas,
Nucl. Phys. B289 (1987) 87.

\noindent --H. Kawai, D. Lewellen, J. Schwartz, H. Tye,
Nucl. Phys. B299 (1988) 431.

\noindent --I. Antoniadis and C. Bachas,
Nucl. Phys. B298 (1988) 586.

\noindent --D. Lewellen, Nucl. Phys. B337 (1990) 61.

\noindent --I. Antoniadis, J. Ellis, J. Hagelin,
and D. Nanopoulos, Phys. Lett. B231 (1989) 65.

\noindent --S. Kalara, J. Lopez, D. Nanopoulos,
Nucl. Phys. B353 (1991) 650.

\noindent --J. Lopez, D. Nanopoulos, and K. Yuan,
Nucl. Phys. B399 (1993) 654.

\noindent --A. Faraggi,
Phys. Lett. B274 (1992) 47;
Phys. Lett. B278 (1992) 131;
Nucl. Phys. B387 (1992) 239;
Phys. Lett. B339 (1994) 223.

\noindent --S. Chaudhuri, S. Chung, G. Hockney,
J. Lykken,
``String consistency for unified model building'',
hepph/9501361, to appear in Nuclear Physics B.

\noindent --K. Dienes, A. Faraggi,
Phys. Rev. Lett. 75 (1995) 2646;
hepth/9505046.

\noindent --S. Chaudhuri, G. Hockney, J. Lykken,
Phys. Rev. Lett. 75 (1995) 2264.

\noindent --S. Chaudhuri, G. Hockney, J. Lykken,
``Three generations in the fermionic construction'',
hepth/9510241.

\vskip .3in
\noindent{\it Supersymmetry Breaking:}
\vskip .3in

\noindent --H. Nilles, Phys. Rep. 110 (1984) 1.

\noindent --S. Ferrara, L. Girardello, H. Nilles,
Phys. Lett. B125 (1983) 457.

\noindent --M. Dine, R. Rohm, N. Seiberg, E. Witten,
Phys. Lett. B156 (1985) 55.

\noindent --L. Dixon, ``Supersymmetry breaking
in string theory'',
Talk at DPF meeting, Rice Univ. 3-6 Jan. 1990.

\noindent --V. Kaplunovsky and J. Louis,
Phys. Lett. B306 (1993) 269.

\noindent --C. Munoz, ``Extracting predictions from supergravity/superstring
for the effective theory below the Planck scale'',
Talk at Beyond the Standard Model IV, hepph/9503314.

\noindent --E. Halyo, ``Supersymmetry breaking
by hidden matter
condensation in superstrings'',
hepph/9505214.

\noindent --Z. Lalak, A. Niemeyer, H. Nilles,
Nucl. Phys. B453 (1995) 100.

\noindent --M. Dine, ``Issues in dynamical
supersymmetry breaking'',
hepth/9503035.

\vskip .3in
\noindent{\it Orbifolds:}
\vskip .3in

\noindent --L. Dixon, D. Friedan, E. Martinec, S. Shenker, Nucl. Phys. B282
(1987) 13.

\noindent --L. Ibanez, H. Nilles, F. Quevedo,
Phys. Lett. B187 (1987) 25.

\noindent --L. Ibanez, J. Kim, H. Nilles, F. Quevedo,
Phys. Lett. B191 (1987) 282.

\noindent --L. Ibanez, J. Mas, H. Nilles, F. Quevedo,
Nucl. Phys. B301 (1988) 157.

\noindent --J. Casas, C. Munoz,
Phys. Lett. B209 (1988) 214;
Phys. Lett. B212 (1988) 343;
Phys. Lett. B214 (1988) 63;
Nucl. Phys. B332 (1990) 189.

\noindent --J. Casas, M. Mondragon, C. Munoz,
Phys. Lett. B230 (1989) 63.

\noindent --A. Font, L. Ibanez, F. Quevedo, A. Sierra,
Nucl. Phys. B337 (1990) 119.

\noindent --L. Ibanez, ``Strings, unification and
dilaton/moduli induced
SUSY-breaking'', Talk at STRINGS '95, hepth/9505098.

\noindent --G. Aldazabal, A. Font, L. Ibanez, A. Uranga,
Nucl. Phys. B452 (1995) 3.

\bye